\documentclass[fleqn,usenatbib]{mnras}
\usepackage[T1]{fontenc}
\usepackage{ae,aecompl}
\usepackage{amstext}
\usepackage{url}
\usepackage[]{times,refname,amsmath,bm}
\bibpunct{(}{)}{;}{a}{}{,}
\usepackage{tabularx}
\usepackage{graphicx,epsfig,color,latexsym}

\renewcommand{\d}{\mathrm{d}}

\newcommand{\bea}{\begin{eqnarray}}
\newcommand{\eea}{\end{eqnarray}}
\newcommand{\be}{\begin{equation}}
\newcommand{\ee}{\end{equation}}
\newcommand{\rund}[1]{\left(#1\right)}
\newcommand{\vc}[1]{\mbox{\boldmath $#1$}}

\newcommand{\eck}[1]{\left[ #1 \right]}

\def\elabel#1{\label{eq:#1}}
\renewcommand{\exp}{\mathrm{exp}}

\title[double-plane lens]{On the double-plane plasma lensing}

\author[Er, Wagner \& Mao]%
       {Xinzhong Er$^1$ \thanks{phioen@163.com},
       	 Jenny Wagner \thanks{thegravitygrinch@gmail.com},
         Shude Mao$^{2,3}$\\
$^1$South-Western Institute for Astronomy Research, Yunnan University, Kunming, P.R.China\\
$^2$Department of Astronomy, Tsinghua University, 100084 Beijing, P.R.China\\
$^3$National Astronomical Observatories, Chinese Academy of Sciences, 20A Datun Road, Chaoyang District, Beijing 100101, P.R.China\\
\\
}
\begin{document}
\maketitle

\begin{abstract}
  Plasma lensing is the refraction of low-frequency electromagnetic rays due to cold free electrons in the universe. For sources at a cosmological distance, there is observational evidence of elongated, complex plasma structures along the line of sight requiring a multi-lens-plane description. To investigate the limits of single-plane plasma lensing, we set up a double-plane lens with a projected Gaussian electron density in each lens plane. We compare double-plane scenarios with corresponding effective single-plane configurations. Our results show how double-plane lenses can be distinguished from single-plane lenses by observables, i.e.~resolved multiple image positions, relative magnifications, time delays, and pulse shapes.
For plasma lensing of fast radio bursts, the observed pulse shape may be dominated by the lensing effect, allowing us to neglect the intrinsic source pulse shape to distinguish different lensing configurations.
The time-domain observables turn out to be the most salient features to tell multi- and single-plane lenses apart.

\end{abstract}

\vspace{1.0\baselineskip}

\section{Introduction}

Gravitational lensing is a powerful tool in modern astrophysics \citep[see e.g.][for reviews]{SEF92,2000rofp.book.....P}. In strong lensing the main lens dominates the lensing effects of matter along the line of sight \citep[e.g.][]{2006glsw.conf...91K,2010CQGra..27w3001B}. But it has been shown that in order to achieve precision cosmology, the line of sight effect has to be taken into account \citep[e.g.][]{2013ApJ...768...39G}. One needs to generalise the lens theory to multiple lens planes to account for perturbing masses along the entire line of sight \citep[e.g.][]{1986ApJ...310..568B,2019A&A...624A..54S}. 
Inhomogeneous distributions of plasma influence the propagation of electro-magnetic signals and cause deflections that can be described in a similar manner as gravitational lensing. Such a phenomenon is therefore called ``plasma lensing''. Different from gravitational lensing, plasma lensing effects depend strongly on the frequency of the signal, and become significant at low frequency, e.g. usually in the radio band. The observation of plasma lensing started decades ago, when abrupt changes were found in the flux density of compact radio sources, which were
attributed to Extreme Scattering Events \citep[ESEs,][]{ESE0}. Although detailed physical models describing all aspects of ESEs remain open, it is widely accepted that plasma lensing effects cannot be neglected in the description of these phenomena. Moreover, millisecond duration pulses known as fast radio bursts \citep[FRBs, e.g.][]{2007Sci...318..777L,2018NatAs...2..842P,frbreview2019,2019A&ARv..27....4P},
may also be subject to plasma lensing in a dense medium, though it is unknown if this environment is located in the host galaxy, or is an intervening structure along the line of sight. 

Additionally it has been observed that the 2-dimensional dynamic power spectra of some pulsars contain remarkably organised parabolic structures
\citep[e.g.][]{stinebring,2007ASPC..365..254S}, which can be explained by highly-anisotropic scattering of radio waves from the pulsar \citep[e.g.][]{2004MNRAS.354...43W,2006ApJ...637..346C}. The radio pulsar time delay has been attributed to plasma structures in the Interstellar Medium (ISM) \citep[e.g.][]{2017MNRAS.464.2075S}. In addition,
\citet{rogers17a} found that plasma distributions around a compact object can compensate the gravitational lensing effect such that the shape of the lens object is altered in its appearance for an observer.

To describe phenomena like ESEs, refractive plasma lensing models have been developed in a similar way as gravitational lensing models. One dimensional Gaussian plasma lensing was initiated to explain ESEs \citep{cleggFL1998}, which has been widely applied in the literature. Since then parametric models of isolated plasma distributions have been put into study, such as the exponential and the power law profiles \citep[e.g.][]{FRBplasma1,er&rogers18,2018MNRAS.478..983S,2018MNRAS.481.2685D,2019PhRvD.100d4006C}. In order to account for the more realistic distribution of the plasma, elliptical or even plasma sheet models have been also proposed \citep[e.g.][]{2012MNRAS.421L.132P,2014MNRAS.442.3338P,2016MNRAS.458.1289L,er&rogers19,2019MNRAS.486.2809G,2019MNRAS.489.3692G}. Case-by-case modelling for the observational events has been proposed, such as a superposition of two 1-dimensional Gaussians for a slice across the plasma density to model the typical U-shape or W-shape light curves
\citep{ESE2017,kerr2018}. Moreover, it has been noticed that magnetic fields are non-negligible as well \citep{2019MNRAS.484.5723L,2020MNRAS.493.1736R}. Since plasma lensing happens usually in low frequency observation, geometric optics may no longer apply and wave effects can be significant \citep{2018arXiv181009058G,2020MNRAS.497.4956J}. The wave effects in multi-plane have also been studied recently \citep{2020arXiv201003089F}. The column density profile and scale of the plasma clumps have also been investigated by modelling independent inversion methods
\citep[e.g.][]{2015ApJ...808..113C,bannister2016,Tuntsov2016}. 

Given the sparse observational constraints, model degeneracy has been found \citep[e.g.][]{Tuntsov2016}. Using the same formalism to describe gravitational lensing and plasma lensing, a general description of all occurring degeneracy in the single-plane plasma lensing formalism was set up in Wagner \& Er (in prep.). Analogous to the freedom to alter the mass density in the lens plane in gravitational lensing, we found a ``gas-sheet degeneracy'' in plasma lensing, allowing one to redistribute the electron density in the plasma lens plane.

Despite detailed studies on both theory and observation, there are lingering difficulties that remain regarding the nature of plasma lensing. For example, it is difficult to interpret the high density and pressures within a single isolated plasma clump. The dispersion measure (DM), i.e. the integrated column density of free electrons between the observer and the source, necessary to account for the lensing effect is too large ($\sim10^3$ cm$^{-3}$) for a structure in the ionised ISM \citep{cleggFL1998}, and cannot exist in pressure balance with the ambient ISM in the Milky Way \citep[e.g.][]{2002astro.ph..7156C}. It is true that currently the model of plasma lensing is simplistic to account for all the aspects. Such an over-simplification has been noticed in gravitational lensing already, e.g. \citet{2020Sci...369.1347M}. Similar ideas have been put forth, e.g., modest 3D electron densities will be required if highly elongated plasma sheets are seen from an edge-on perspective \citep{romani87,2010ApJ...708..232B,2014MNRAS.442.3338P,plasmaSheets,2018MNRAS.478..983S}. Plasma sheets have also been found in numerical simulations of the supernova-driven turbulence \citep[e.g.][]{2012ApJ...750..104H}. 

In gravitational lensing, a single lens plane is often sufficient for an effective and precise description of the observations, as mentioned earlier, because there are only very few strongly deflecting, separate mass agglomerations that are highly aligned along the line of sight. In contrast, the deflecting structures producing plasma lensing are diffuse ion clouds that can be extended along the line of sight such that multiple lens planes may be required to model the observations. For example, it has been suggested of multiple plasma sheets in pulsar observation \citep[e.g.][]{2006ChJAS...6b.233P}. More than one plasma screen have been identified along the way to the central black hole \citep{2017MNRAS.471.3563D}. Moreover, it is widely accepted that the observed DM of FRBs have various contributors: the host galaxy, inter-galactic medium (IGM) and the Milky Way etc \citep[e.g.][]{thornton2013,FRBplasma1,yang2017,2019A&ARv..27....4P,2020Natur.581..391M}. It has been proposed to use FRBs to study the missing baryons in the universe \citep[e.g.][]{2020Natur.581..391M,2014ApJ...780L..33M,2021MNRAS.503.4576D,2021arXiv210108005J}. Thus, in order to precisely count the electrons in the IGM, it is necessary to determine the locations of free electron clouds along the line of sight. 
For a source at a cosmological distance, the lens distance can cause dramatic changes of the lensing effects. It is possible to model the system by a multi-plane lens. 
We will summarise the thin lens plasma lensing formalism, especially the Gaussian model in Sect.\,\ref{sec:basic} and introduce the multi-plane approximation of plasma lensing in Sect.\,\ref{sec:multiplanes} and compare it with an effective description by a single thin lens in Sect.\,\ref{sec:simulation}. We summarise our conclusions in Sect.\,\ref{sec:conclusions}. In this paper we adopt the standard $\Lambda$CDM cosmology with parameters based on the results from the $Planck$ data \citep{2018arXiv180706209P}: $\Omega_{\Lambda}=0.6847$, $\Omega_\mathrm{m}=0.3153$, and Hubble constant $H_0=100h$\,km\,s$^{-1}$\,Mpc$^{-1}$ with $h=0.6736$.

\section{The thin lens formalism}
\label{sec:basic}
We first outline the basic formalism for the single-lens-plane plasma lensing. More details can be found in \citet{er&rogers18}. 
The notation mainly follows the general reviews in gravitational lensing \citep{SEF92, 1996astro.ph..6001N}. We denote the angular diameter distances between the source and the lens as $D_{\rm ds}$, between the source and
the observer as $D_{\rm s}$ and between the lens and the observer as $D_{\rm d}$. We introduce the angular coordinates $\vc{\theta}=(\theta_x,\theta_y)$, which are perpendicular to the line of sight, and those in the source plane as $\vc{\beta}=(\beta_x,\beta_y)$. The subscripts $x,y$ stand for the two directions on the sky. The coordinates in the lens and source planes are related through the lens equation
\be
\vc{\beta} = \vc{\theta} - \vc{\alpha(\theta)} = \vc{\theta} - \nabla_{\vc{\theta}} \psi(\vc{\theta}) \;,
\elabel{lenseq}
\ee
where $\vc{\alpha(\theta)}$ is the deflection angle, $\psi(\vc{\theta})$ is the effective lens potential and $\nabla_{\vc{\theta}}$ is the gradient in the lens plane with respect to $\vc{\theta}$. 

The refractive index of a cold plasma for light with angular frequency $\omega=2\pi \nu$ is given by
\be
n^2_{\rm pl} \equiv 1- \omega^2_p/\omega^2 \;,
\ee
where $\omega^2_\mathrm{p}\equiv e^2 n_\mathrm{e}/(\epsilon_0 m_\mathrm{e})$ is the plasma frequency, $e$ is the electron charge, $m_\mathrm{e}$ is the mass of the electron, $\epsilon_0$ is the vacuum permittivity, and $n_\mathrm{e}$ is the 3-dimensional number density of electrons in the plasma. The deflection caused by a clump of plasma mainly comes from the gradient of the electron density \citep{2010MNRAS.404.1790B}. If the plasma frequency is much smaller than the observational frequency $\omega_\mathrm{p} \ll \omega$, the deflection angle is given by
\be
\hat\alpha (b) = -\frac12 b\int \frac{\omega_\mathrm{p}^2}{\omega^2} \frac{1}{r\, n_\mathrm{e}(r)}
\frac{\d n_\mathrm{e}(r)}{\d r} \d l \;,
\ee
where $r$ is the 3-dimensional coordinate, $b$ is the impact parameter and $l$ is the coordinate along the line of sight. 
The reduced deflection angle in the lens equation is given by $\alpha = \hat{\alpha}D_{\rm ds}/D_{\rm s}$. In the thin lens approximation, we project the electron distribution $n_\mathrm{e}$ on the lens plane to obtain a 2-dimensional surface density profile
\be
N_\mathrm{e}(\vc{\theta}) \equiv \int_0^{D_{\rm s}}  n_\mathrm{e}(\vc{\theta}, l) \d l \;.
\ee
Similar as in gravitational lensing, the deflection angles usually are small.
The integrals can thus be done along unperturbed rays, and $N_\mathrm{e}$ is approximated by the projected electron density at the image position. 

The propagation of the signal in the medium is given by the observable group velocity, and will be delayed. The time delay with respect to a signal propagating through vacuum can be approximated as
\be
T_{\rm DM} \approx \frac{(1+z_\mathrm{d})}{2c} \int_0^r \frac{\omega^2_\mathrm{p}}{\omega^2} \d l
= \frac{1}{(1+z_\mathrm{d})} \frac{\lambda^2 r_\mathrm{e}}{2\pi c}\,N_\mathrm{e}(\vc{\theta}) \;,
\ee
where $\lambda=c/\nu$ is the observed wavelength of the photon\footnote{$\lambda=(1+z_{\rm d})\lambda_{d}$, where $\lambda_{d}$ is the wavelength of a photon at the redshift of the lens $z_\mathrm{d}$. Such effect merges with the pre-factor of cosmological time dilation and leaves $1/(1+z_{\rm d})$ in the equation \citep[e.g.][]{2003ApJ...598L..79I,2004MNRAS.348..999I}. Since we study lenses at low redshift in this work, we neglect such a difference.}. 
Both the time delay and deflection of signals in plasma depend on the wavelength.
We first consider cases of a single wavelength, so $\lambda$ is held fixed until stated otherwise.
The classical electron radius is given by $r_\mathrm{e}=e^2/(4\pi \epsilon_0 m_\mathrm{e} c^2)$.
Under these prerequisites, the pulsar Dispersion Measure (DM) gives a similar projected density and arrival time difference. 
Yet, such an approximation is only valid under the condition that the geometric time delay caused by the deflection angle is small compared with the dispersive delay caused by the DM \citep{er+2020}.

The total time delay as a combination of the dispersive delay and the geometric delay gives the difference in travel time between the lensed ray and an unperturbed ray propagating in the vacuum background cosmology
\be
T(\vc{\theta},\vc{\beta})= \frac{(1+z_\mathrm{d})}{c}\frac{D_{\rm d} D_{\rm s}}{D_{\rm ds}}
\eck{\frac{(\vc{\beta}-\vc{\theta})^2}{2} + \frac{1}{(1+z_{\rm d})^2} \psi(\vc{\theta})} \;,
\elabel{timedelay}
\ee
where we define the ``effective plasma lens potential'' in a similar fashion to gravitational lensing by
\be
\psi(\vc{\theta}) \equiv
\dfrac{D_{\rm ds}}{D_{\rm d} D_{\rm s}} \frac{\lambda^2}{2\pi} r_\mathrm{e} N_\mathrm{e}(\vc{\theta}) \;.
\elabel{psi}
\ee
The deflection angle by plasma lensing can be calculated from 
\be
\alpha=\nabla_{\theta} \psi(\vc{\theta}),
\ee
where $\nabla_{\theta}$ is the gradient on the image plane.

Usually, geometrical optics is an excellent approximation in lensing. Wave effects can be important for long wavelength observations and coherent emissions of the sources, when scattering structures in the plasma have small scales, see e.g.~\citet[][]{2018arXiv181009058G,2020MNRAS.497.4956J,2020arXiv201003089F,2021arXiv211007119J}. In \citet{2014MNRAS.442.3338P}, a rough estimate has been proposed, i.e. for the structures smaller than the Fresnel scale
\be
r_{\rm F}\equiv \sqrt{\lambda D_d}\sim 0.03 {\rm AU} \rund{\frac{\lambda}{m}}\rund{\frac{D_d}{\rm kpc}},
\ee
wave optics is necessary to describe the lensing deflection. For example, for observations at 1 GHz, a plasma structure of a few tens AU in IGM has the possibility to cause wave effects.
Moreover, when the source is close to the caustics, wave effects can be significant even when the lens is not small compared to the wavelength. Such events will, however, produce extremely large magnifications and probably have not been observed or identified yet. Thus, we limit our study to the geometrical optics limit, and to the cases when the sources are not extremely close to the caustics.
\subsection{Multi-plane formalism}
\label{sec:multiplanes}
\begin{figure}
	\centerline{\includegraphics[width=10.0cm]{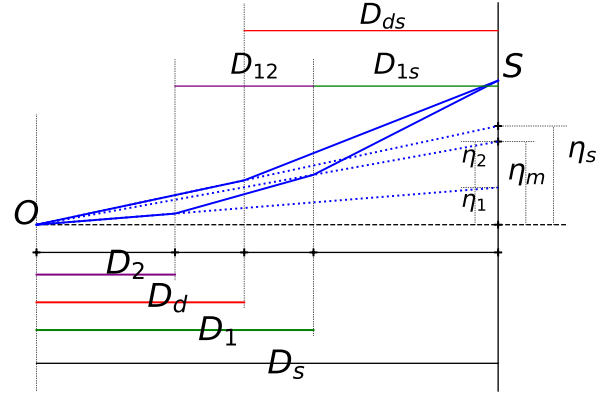}}
	\caption{Comparison between a single-plane lens at a distance $D_\mathrm{d}$ from the observer and a double-plane lens with a foreground lens at distance $D_2$ from the observer and a background lens at distance $D_1$ from the observer. The physical image position in the source plane is $\eta=\theta D_s$.}
	\label{fig:multi-plane}
\end{figure}

We now consider a lens system with multiple plasma clouds between the observer and the source. Using the formalism described in the previous section, we adopt the thin lens approximation in each lens plane.
For each of $n$ lens planes, Eq.\,\ref{eq:lenseq} then becomes
\be
\vc{\beta}_i= \vc{\theta}_i-{\cal D}_i \vc{\alpha}_i(\vc{\theta}_i) \;, 
\ee
where the subscript $i$ stands for the $i^{\rm th}$ lens \citep{1993A&A...268..453E}. 
The scaling factor of each lens plane to the final source plane is 
\be
{\cal D}_i=\dfrac{D_{i\,(i-1)}D_{\rm s}}{D_{i{\rm s}}D_{i-1}} \;,
\ee
where $D_i$ is the $i^{\rm th}$ lens distance, and $D_{i\,(i-1)}$ is the distance between the $i^{\rm th}$ and $(i-1)^{\rm th}$ lens plane. 
Assuming the lens is centred at the origin, the corresponding deflection angle is
\be
\vc{\alpha}_i(\vc{\theta}_i) = - \dfrac{\lambda^2 r_\mathrm{e} N_{0i}}{2 \pi}\frac{D_{i{\rm s}}}{D_{\rm s} D_{i}}
\frac{\vc{\theta}_i}{\sigma_i^2} \, {\rm exp}\rund{-\frac{\vc{\theta}_i^2}{2\sigma_i^2}} \;.
\elabel{multialpha}
\ee
The image position from the previous lens plane will be the source position for the next lens plane, i.e. $\vc{\beta}_i=\vc{\theta}_{i-1}-\vc{c}_i$, where $\vc{c}_i$ is the origin of the $i^{\rm th}$ lens. 
Fig.~\ref{fig:multi-plane} visualises the configuration of a double plane lens.
Then, the complete lens equation can be written as
\be
\vc{\beta} = \vc{\theta} -\sum_{i=0}^n \vc{\alpha}_i(\vc{\theta}_i) + \sum_{i=0}^n \vc{c}_i \;,
\elabel{multilenseq}
\ee
where $\vc{\beta}$ is the initial source position, $\vc{\theta}$ is the final image position. The scaling ${\cal D}_i$ becomes unity for the lens plane closest to the source. 

The lens structures along the line of sight can be complicated. The fluctuations in the electron density may follow a Kolmogorov spectrum \citep[e.g.][]{1985ApJ...288..221C}, and can have variations on scales down to sub-AU. Structures on different scales can cause various strengths of deflections depending on the lens distances and offsets with respect to the angular position of the source. We focus on those electron density perturbations that cause macro lensing effects, i.e. the small fluctuations which cause scattering of the image and wave effects will not be considered in this work \citep[e.g.][]{1989MNRAS.238..963N}.

The magnification $\mu$ of an observed image generated by a multi-plane lens can be calculated from the Jacobian matrix $A$ of its source position $\vc{\beta}$ and its final image position $\vc{\theta}$.
Its entries are given by 
\begin{align}
A_{pq}\rund{\vc{\theta}} &= \rund{\frac{\partial {\beta_p}}{\partial {\theta_q}}} \nonumber\\
&= \delta_{pq} - \sum \limits_{i=0}^n \sum\limits_{r=x,y} \frac{\partial \alpha_{ip}\rund{\vc{\theta_i}}}{\partial \theta_{ir}} \frac{\mathrm{d} \theta_{ir}}{\mathrm{d} \theta_q}\;, \quad p,q=x,y \;,
\label{eq:distortion_matrix}
\end{align}
such that $\mu = \det{(A)}^{-1}$.
Here, we introduced $p,q$ as indices for the $x$- and $y$-directions, while $i$ denotes the index counting the lens planes.

The time delay in a multi-plane lens with respect to an undeflected path is given by
\begin{align}
T(\vc{\theta}) &= \sum_{i=0}^{n} \frac{1+z_{{\rm d} i}}{c} \nonumber\\
&\eck{\frac{D_{i} D_{i-1}}{ D_{i\,(i-1)}}\frac{(\vc{\beta}_i -\vc{\theta}_i)^2}{2} + \frac{1}{(1+z_{{\rm d}i})^2}\frac{\lambda^2 r_\mathrm{e}}{2 \pi}N_{{\rm e} i}(\vc{\theta}_i)} \;.
\end{align}
It can also be written in terms of the lens potential using Eq.\,\ref{eq:psi} for all lens planes
\be
T(\vc{\theta})=\sum_{i=0}^{n}  \frac{1+z_{{\rm d}i}}{c}\frac{D_{i} D_{i-1}}{D_{i\,(i-1)}}
\eck{\frac{(\vc{\beta}_i-\vc{\theta}_i)^2}{2} + \frac{1}{(1+z_{{\rm d}i})^2} \psi_i(\vc{\theta}_i)} \;.
\label{eq:multidelay}
\ee


\section{Lens models}
\label{sec:lens_models}

\subsection{Axisymmetric Gaussian lens}
\label{sec:Gauss_lens}

The axisymmetric Gaussian density distribution as plasma lens was introduced by \citet{cleggFL1998} to model observations of the extra-galactic sources 0954+654 and 1741-038.
It provides a good statistical description to the stacked electron distribution. Moreover, Gaussian plasma clumps can serve as building blocks for a more complex symmetric lens configuration. We thus adopt a Gaussian model as our example lens.
Employing polar coordinates with radius $\theta > 0$, we specify the projected Gaussian electron distribution on the lens plane
\be
N_\mathrm{e}(\theta)=N_0\,{\rm exp}\rund{-\frac{\theta^2}{2\sigma^2}}\;,
\label{eq:Ne}
\ee
with $N_0$ as the maximum electron column density, i.e. $N_\mathrm{e}(\theta=0)$,
and $\sigma$ as the width of the clump. 
Inserting Eq.~\ref{eq:Ne} into Eq.~\ref{eq:psi} potential yields the radially symmetric
\be
\psi(\theta)= \theta_0^2 \exp\left( -\frac{\theta^2}{2\sigma^2} \right)
\label{eq:exp-pot}
\ee
and deflection angle
\be
\alpha(\theta)=-\theta_\text{0}^2
\frac{\theta}{\sigma^2}\exp\left( -\frac{\theta^2}{2 \sigma^2} \right) \;,
\label{deflExp}
\ee
where the characteristic angular scale is given by
\be
\theta_0 = \lambda \left(\frac{D_\text{ds}}{D_\text{s} D_\text{d}}
\frac{1}{2\pi} r_\text{e} N_\text{0} \right)^\frac{1}{2}.
\elabel{theta0}
\ee
Together with $\sigma$, one can determine the area of lensing magnification. It is the analogue of the Einstein radius in gravitational lensing.

\subsection{Combined Gaussian lenses}
\label{sec:combined_lens}

Putting two Gaussian lenses a distance $\overline{\vc{\theta}}_1$ apart from each other in the same lens plane, the total deflection potential reads
\begin{align}
\psi_{\rm b}(\vc{\theta}) &= \theta^2_{01}\,{\rm exp} \rund{-\dfrac{(\vc{\theta}-\bar{\vc{\theta}}_1)^2}{2\sigma_1^2}} + \psi_2(\vc{\theta}),
\label{eq:binary_potential}
\end{align}
where 
\be
\psi_2(\vc{\theta})=\theta^2_{02}\,{\rm exp}\rund{-\dfrac{\vc{\theta}^2}{2\sigma_2^2}}
\label{eq:psi2}
\ee
and $\theta_{01}$ and $\theta_{02}$ are given by Eq.~\ref{eq:theta0} with the same distances between the lens, source, and observer, but they can have different electron densities $N_{01}$ and $N_{02}$, and different lens scale $\sigma_1$ and $\sigma_2$. 
Without loss of generality, the angular position of the second lens is at the origin of the coordinate system and the first lens is located at distance $\bar{\vc{\theta}}_1$ along the $x$-axis, i.~e., $\bar{\vc{\theta}}_1 = (\bar \theta_1,0)$. 
This lens is called a 2D binary lens (with subscript b). 

Further generalising the lens model, we can also put the two Gaussian lenses of the previous model at two different distances along the line of sight to obtain the deflection potential of a 3D double lens (with subscript d)
\begin{align}
    \psi_{\rm d}(\vc{\theta}) &= \theta^2_{01}\,{\rm exp}\rund{-\dfrac{(\vc{\theta}(1+x)-\bar{\vc{\theta}}_1)^2}{2\sigma_1^2}} + \psi_2(\vc{\theta})\;,
    \label{eq:double_potential}
\end{align}
where the factor $x$ accounts for the deflection effects due to the distance between the two lens planes, and will be given in the next subsection (Eq.~\ref{eq:ks}).
The two Gaussian lenses are now located at distances $D_1$ and $D_2$ from the observer with $D_2 < D_1$. $\psi_2(\theta)$ is the lens potential of the lens at $D_2$ given by Eq.~\ref{eq:psi2}.

\subsection{Analytical comparisons}
\label{sec:analytical_treatment}

To gain an intuition for the results obtained in the simulations of Section~\ref{sec:simulation}, we investigate qualitatively when a binary or double lens may be mistaken as a single Gaussian lens and when a double lens may be mistaken as a binary lens. 

Comparing Eqs.~\ref{eq:exp-pot} and \ref{eq:binary_potential}, it is easy to show that both potentials can only become equal for trivial limits: 
when $N_0 \rightarrow 0$ or $\sigma \rightarrow 0$ for one of the Gaussians, Eq.~\ref{eq:binary_potential} simplifies to Eq.~\ref{eq:exp-pot}.  
If both lenses in the binary deflection potential are so far apart that we only probe one of them, Eq.~\ref{eq:binary_potential} will also become Eq.~\ref{eq:exp-pot} in the limit of $\bar{\theta}_1 \rightarrow \infty$, or both lenses have the same width $\sigma_1 = \sigma_2$ and are both located at the origin in the limit of $\bar{\theta}_1 \rightarrow 0$. 
Thus, these two lens models are distinct and cannot be confused with each other and we can analogously extend this argument to the double lens. 
Yet, as shown in \citep{er&rogers19}, it is possible to model a double lens with one single \emph{elliptical} Gaussian lens for a system with small angular separation. 
In Section~\ref{sec:positions}, we will see that the time delays between the images can distinguish between a single-plane and a multi-plane lens. 

To investigate possible confusions between a binary and a double lens, we express the parameters of the background lens in the double lens in terms of the parameters of the foreground lens
\be
N_{01} = k_\mathrm{N} N_{02} \;, \quad \sigma_1 = k_{\rm \sigma} \sigma_2 \;.
\label{eq:parameters}
\ee
Notice that $k_\mathrm{N}$ and $k_{\rm \sigma}$ may not be small numbers, thus this is a general reformulation without assuming that the background lens is merely a perturber to the foreground lens. 
We further introduce the following abbreviations
\be
k_\mathrm{D} \equiv \frac{D_{1 {\rm s}}}{D_{1}}\frac{D_{2}} {D_{2{\rm s}}} \;, \quad 
x \equiv \frac{D_{21}}{D_{1}D_{2}} \frac{\lambda^2 r_\mathrm{e}}{2\pi} \frac{N_{02}}{\sigma_2^2} {\exp}\rund{-\frac{\vc{\theta}^2}{2\sigma_2^2}} \;.
\label{eq:ks}
\ee
Using these expressions and the notation introduced in Section~\ref{sec:multiplanes}, the potential for a double lens (Eq.\,\ref{eq:double_potential}) can be written as
\begin{align}
\psi_\mathrm{d}\rund{\vc{\theta}} = \psi_2\rund{\vc{\theta}}\Bigg\{ & 1+ k_\mathrm{D} k_\mathrm{N} {\rm exp}\rund{-\frac{\vc{\theta}^2}{2\sigma_2^2}}^{\rund{\frac{(1+x)^2}{k_{\rm \sigma}^2}-1}} \times \nonumber \\ 
& {\rm exp}\rund{-\frac{\overline{\vc{\theta}}_{1}^2 - 2(1+x)\vc{\theta} \cdot \overline{\vc{\theta}}_{1}}{2k_{\rm \sigma}^2 \sigma_2^2}} \Bigg\}\;,
\label{eq:psi_d}
\end{align}
with the scalar product between $\vc{\theta}$ and $\overline{\vc{\theta}}_1$ given as the standard dot-product. It can be seen that $k_\mathrm{N}$ is degenerate with $k_\mathrm{D}$, which is to be expected when one only measures integrated electron densities.
Thus, additional constraints are necessary to break this degeneracy between the line-of-sight distance effects and the total deflecting central electron density when modelling observables with a double lens configuration.

The double lens can be simplified to a binary lens by setting
\be
D_{1} = D_{2} \quad \Rightarrow \quad k_\mathrm{D} = 1 \;, \quad x = 0 \;,
\ee
such that Eq.~\eqref{eq:psi_d} reduces to
\begin{align}
\psi_\mathrm{b}\rund{\vc{\theta}} = \psi_2\rund{\vc{\theta}} \Bigg\{ & 1+ k_\mathrm{N} \exp\rund{-\frac{\vc{\theta}^2}{2\sigma_2^2}}^{(k_{\rm \sigma}^{-2}-1)} \times \nonumber \\ 
& {\exp}\rund{-\dfrac{\overline{\vc{\theta}}_{1}^2-2\vc{\theta} \cdot \overline{\vc{\theta}}_{1}}{2 k_{\rm \sigma}^2 \sigma_2^2}} \Bigg\}.
\label{eq:psi_b}
\end{align}
Hence, one trivial limiting case to convert a double lens into a binary one is the case in which the lens planes are close to each other. 
Comparing Eq.~\ref{eq:psi_d} with Eq.~\ref{eq:psi_b} (or Eq.\,\ref{eq:binary_potential} with Eq.\,\ref{eq:double_potential}), we find that there is another option, namely transforming the parameters of the double lens into those of a binary lens by applying
\be
k_\mathrm{Nb} = k_\mathrm{Nd} k_\mathrm{D} \;, \quad k_\mathrm{\sigma b} = \frac{k_\mathrm{\sigma d}}{1+x} \;, \quad \overline{\vc\theta}_{\rm 1 b} =  \frac{\overline{\vc\theta}_{\rm 1 d}}{1+x} \;. 
\label{eq:double2binary}
\ee
While this transformation allows for a degeneracy between the double and binary lens potential, this degeneracy is broken for the deflection angles because $x$ depends on $\vc\theta$. Thus, even with a high degree of fine-tuning, we cannot easily find a transformation between these two lens models that transforms all lens properties correctly. 
This is to be expected as the general double lens model has more degrees of freedom than the binary lens. 
While we cannot globally transform the double-lens model into a binary one, we can find multiple-image configurations that resemble each other in terms of their image positions and magnifications, as we will show in Section~\ref{sec:simulation}. Yet, these numerical examples also show that the time delays can distinguish the single-plane from the multi-plane model. 

\section{Simulations}
\label{sec:simulation}

\subsection{Lensing deflection and magnification of single-, and double-lenses}
\label{sec:positions}
\begin{table}
  \begin{tabular}{c|c|c c}
parameter &single    & \multicolumn{2}{c}{double}  \\
              &      &D1    &D2\\ 
\hline
$N_0$ (pc\,cm$^{-3}$)    &90     &70   &40 \\
\hline
$z_d$    &0.001  &0.0015  &0.0005\\
\hline
$D_d$ (Mpc)   &4.4  &6.7   &2.2  \\
\hline
$\sigma$ (mas) &2.2  &10   &2.2 \\
\hline
$\theta_0$ (mas) &5.9  &4.2  &5.5\\
\hline
\end{tabular}
\caption{Lens parameters for the single and double lenses of Section~\ref{sec:positions} aligned along the line of sight. For the single lens we use $\sigma=10^4$ AU, for the double lens we use $\sigma=7,0.5\times10^4$ AU. The double lens is a combination of D1 and D2 lenses. }
\label{tab:lens-para2}
\end{table}

\begin{table*}
    \begin{tabular}{c||c|c|c||c|c|c||c|c|c}
    parameter & \multicolumn{3}{c}{single}  & \multicolumn{3}{c}{double}  & \multicolumn{3}{c}{single D2}  \\
    \hline
     image-$i$ & $\theta_i$(mas) &$\mu_i$ &$\Delta t_i$ (ms) & $\theta_i$(mas) &$\mu_i$ &$\Delta t_i$ (ms)  & $\theta_i$(mas) &$\mu_i$ &$\Delta t_i$ (ms)\\
     \hline
     1 &1.56 &0.042 &680 &1.58 &0.044 &528 &1.89 &0.086 &295\\
     2 &3.77 &0.18 &301 &3.92 &0.22 &341 &3.36 &0.22 &173\\
     3 &10.0 &1.0 &0.017 &9.0 &0.9 &209  &10.0  &1.0 &0.007\\
     \hline
    \end{tabular}
    \caption{Comparison between single plane (left), double lens (middle) and D2 lens (right): image position $\theta_i$, magnification $\mu_i$ and time delay $\Delta t_i$ with respect to the unperturbed ray through vacuum. The source position is $\beta=10$ mas. }
    \label{tab:single_vs_dB}
\end{table*}

We compare the image positions and magnifications of lensed images between a single lens and lenses at double planes. 
We first investigate the lensing effect with two individual Gaussian profiles perfectly aligned along the line of sight, thus employ polar coordinates due to the axisymmetry of these lensing configurations. 
To set up our double lenses simulations, we use a background radio source located at a cosmological distance, which is a repeating FRB \citep{repeatingFRB}, $z_s=0.19$ ($\sim$ 690 Mpc), and we only consider a point source, such as an FRB or a pulsar. 
Given this source, the IGM along the line of sight may contribute a significant electron density as well, but it can be distributed in a large redshift range. We compare a single lens case ($z_d=0.001$) and a double lens case ($z_d=0.0005,0.0015$).
Unless stated otherwise, a single observation frequency of $\nu=1$ GHz is used for all the tests.

\begin{figure}
\centerline{\includegraphics[width=7.5cm]{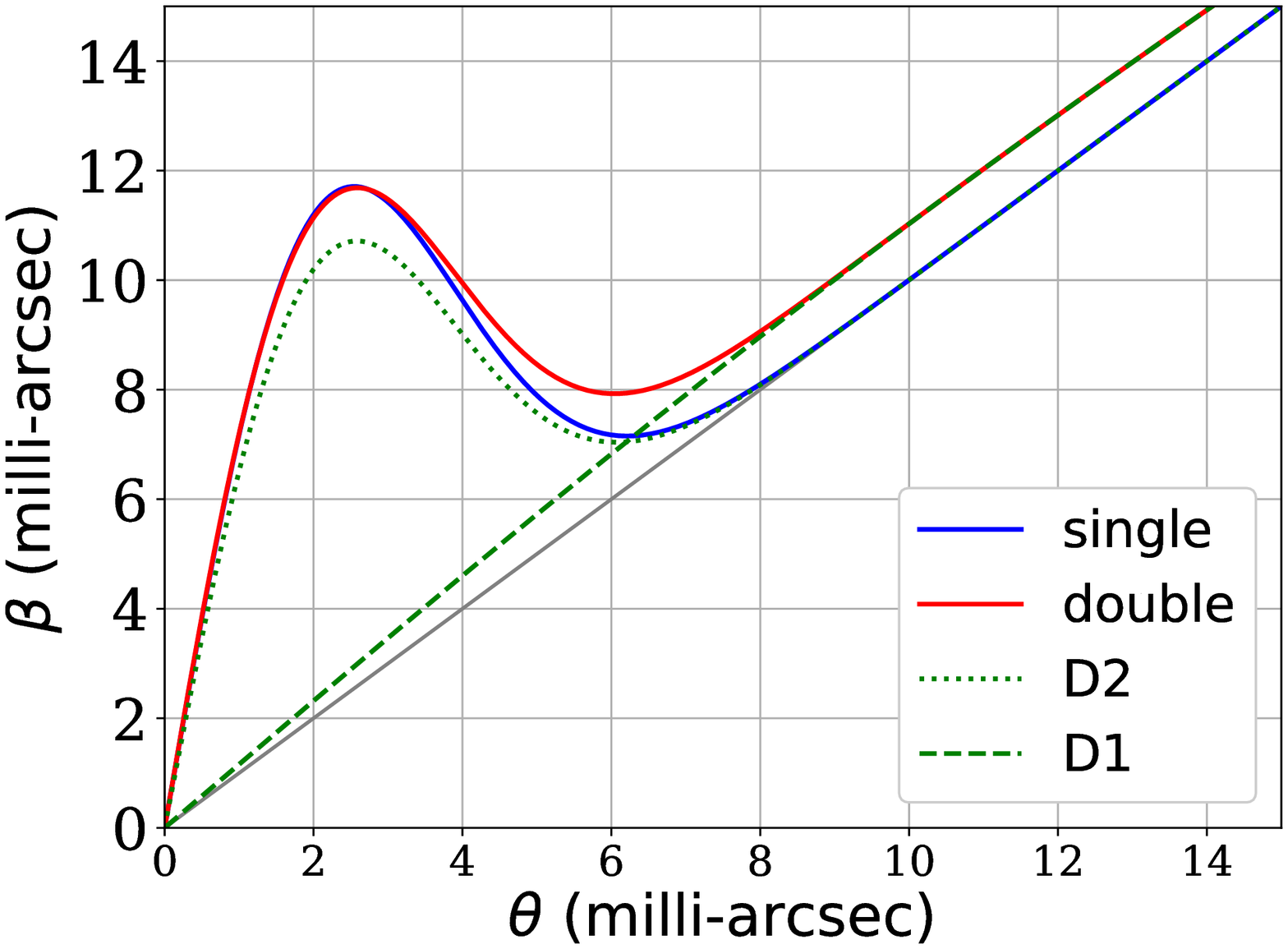}}
\centerline{\includegraphics[width=7.5cm]{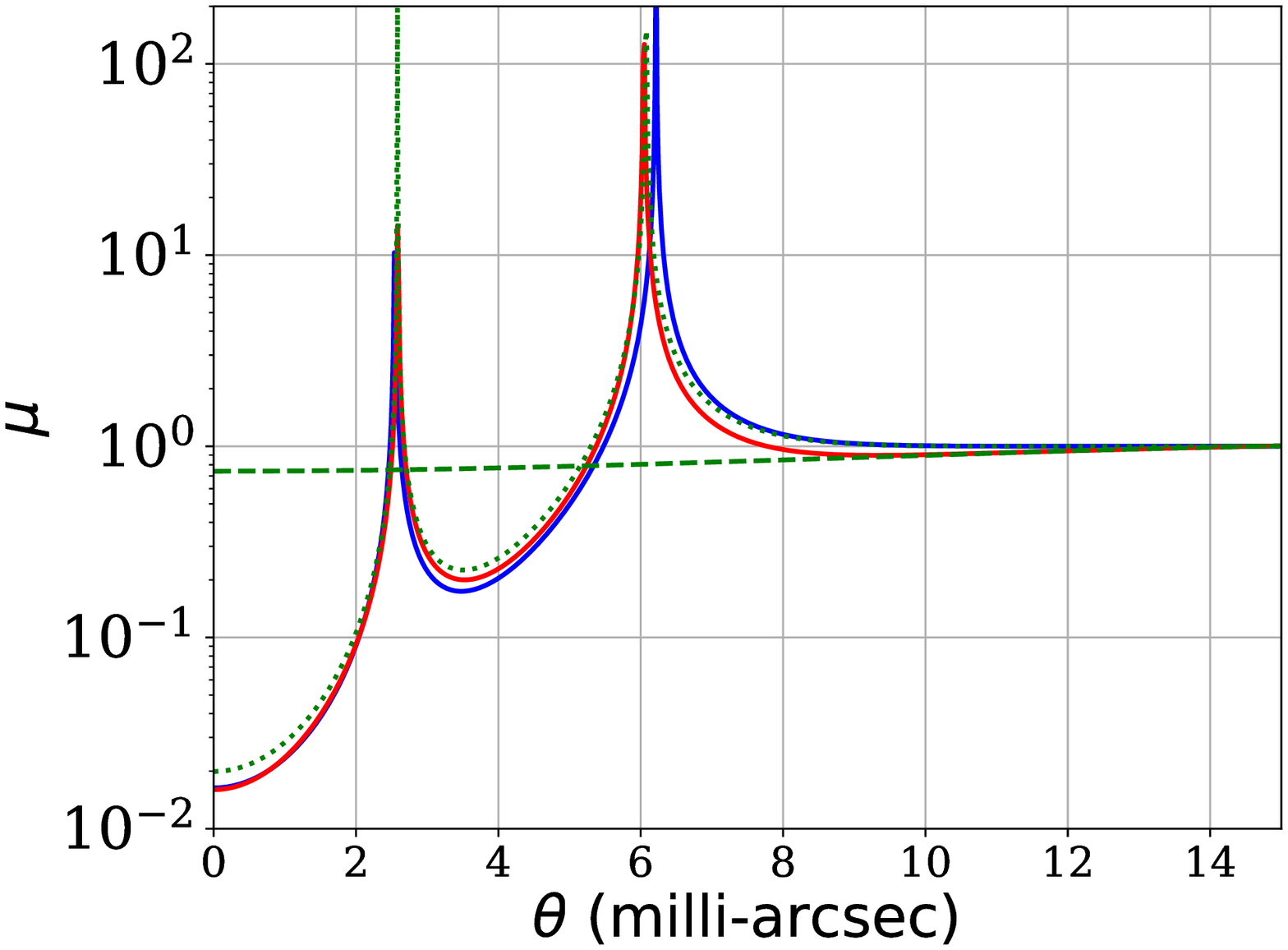}}
\caption{Comparison of double plane lenses with a single plane lens. Top (bottom) panel: image-source position (magnification) curves. See Tables\,\ref{tab:lens-para2} for more details about the lens parameters.}
\label{fig:2lenses}
\end{figure}

Table~\ref{tab:lens-para2} summarises the properties of the single and the double lenses. The multiple image positions $\theta$ and source positions, and magnification curves are plotted in Fig.~\ref{fig:2lenses}. 
The solid red line represents the double lens, which is designed to mimic the single lens (blue solid line) and therefore shows a similar curve in the two plots. One can see that the red curve almost overlaps with the blue one at small $\theta$, i.e. the strong deflection region.
The green curves show the individual contributions of the two single lenses D1 and D2 of double lens. 
The electron density distribution in D1 is relatively smooth. Thus, D1 weakly boosts the deflection of D2 and only dominates the total deflection at large $\theta$, which is relatively weak and will not cause significant magnification.

The single lens, double lens, and D2 lens give similar image position patterns and magnifications which will be difficult to distinguish. In Table~\ref{tab:single_vs_dB}, we show an example of multiple images of the single lens and the double lens. The time delays between image pairs are significantly different for the different lensing configurations, which offers a way to break the degeneracy between the scenarios.
We also note that the magnification of the third image is smaller than 1 in the double-lens configuration. Thus, modelling an observed image configuration caused by this double lens as one of the single-lens configurations shown in Table~\ref{tab:single_vs_dB} will overestimate the magnification of the third image. Consequently, the lens parameters will be biased accordingly.

\subsubsection{3D double lenses and 2D binary lenses}
\label{sec:caustics}
\begin{table}
  \begin{tabular}{c|c c|c c}
parameter & \multicolumn{2}{c}{2D Binary} & \multicolumn{2}{c}{3D Double}  \\
        &B1  &B2      &D1    &D2\\ 
\hline
$z_d$    &0.0015 &0.0015  &0.0015  &0.0005\\
\hline
$\sigma$ (mas) &1.5  &4.5   &1.5   &4.5 \\
\hline
$\theta_0$ (mas) &4.8  &8.3   &4.8  &8.3\\
\hline
\end{tabular}
\caption{Lens parameters for the 2D binary and 3D double lenses of Section~\ref{sec:caustics}.}
\label{tab:lens-offset-para}
\end{table}

\begin{figure*}
\centerline{
\includegraphics[width=4cm]{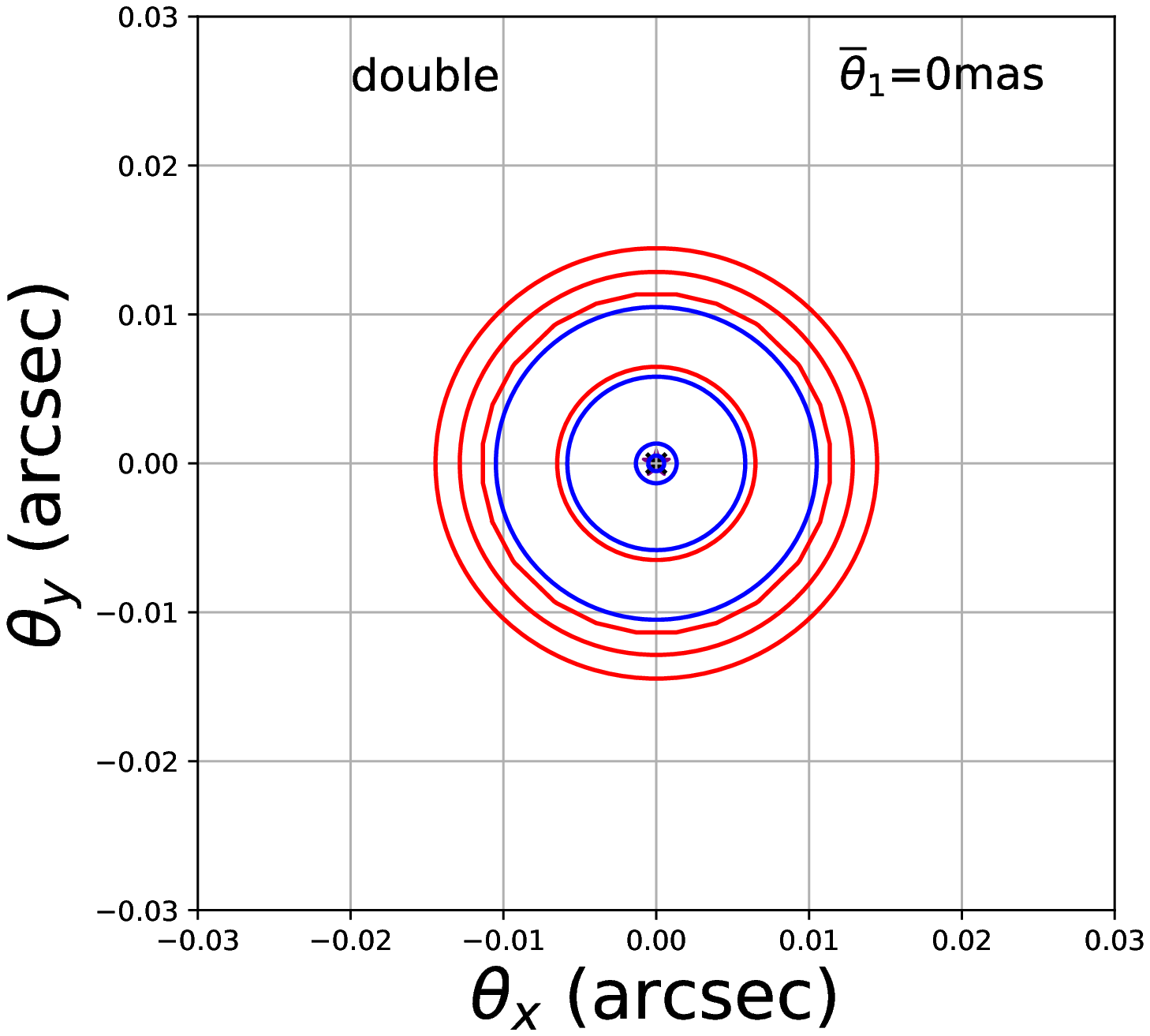}
\includegraphics[width=4cm]{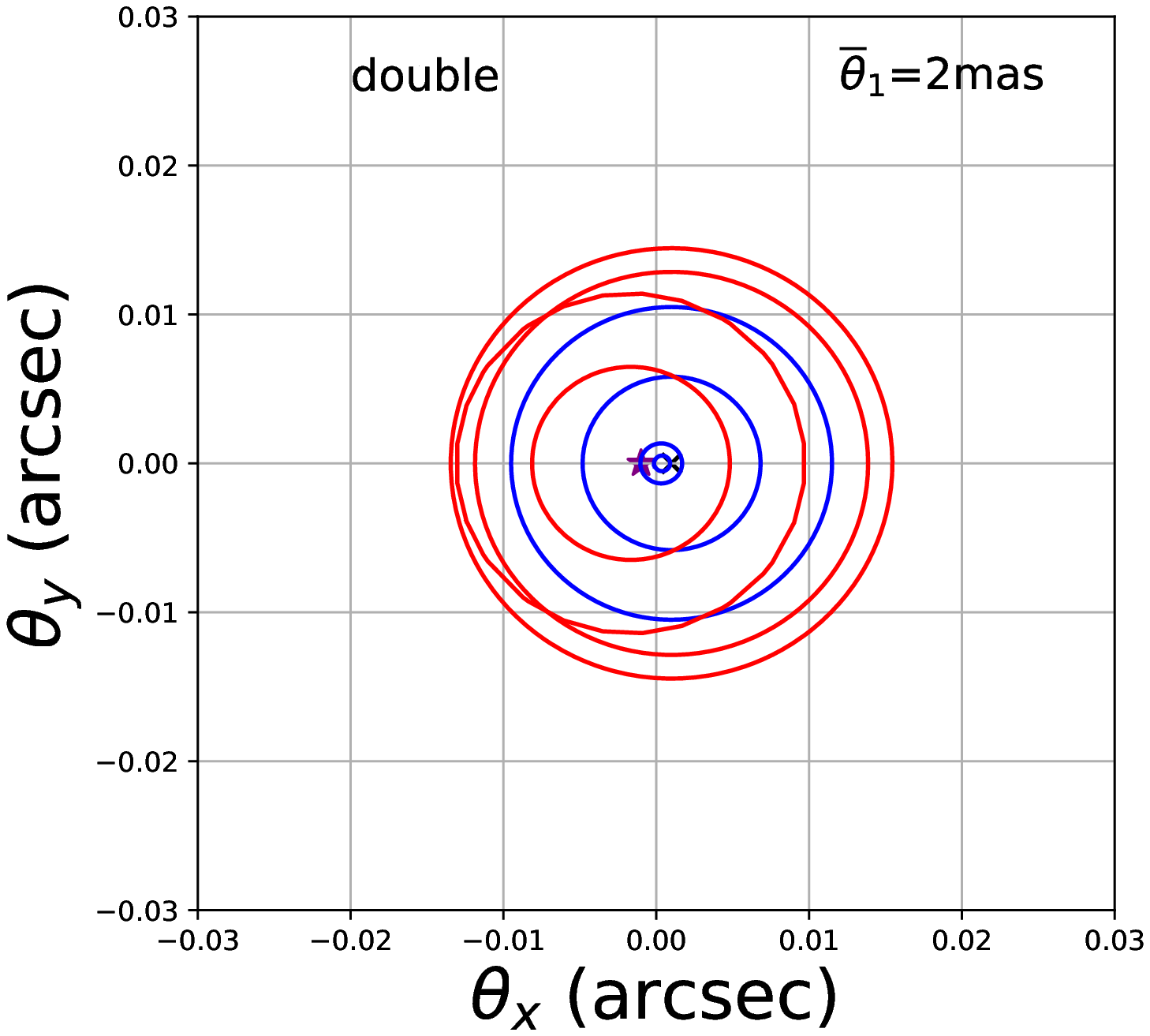}
\includegraphics[width=4cm]{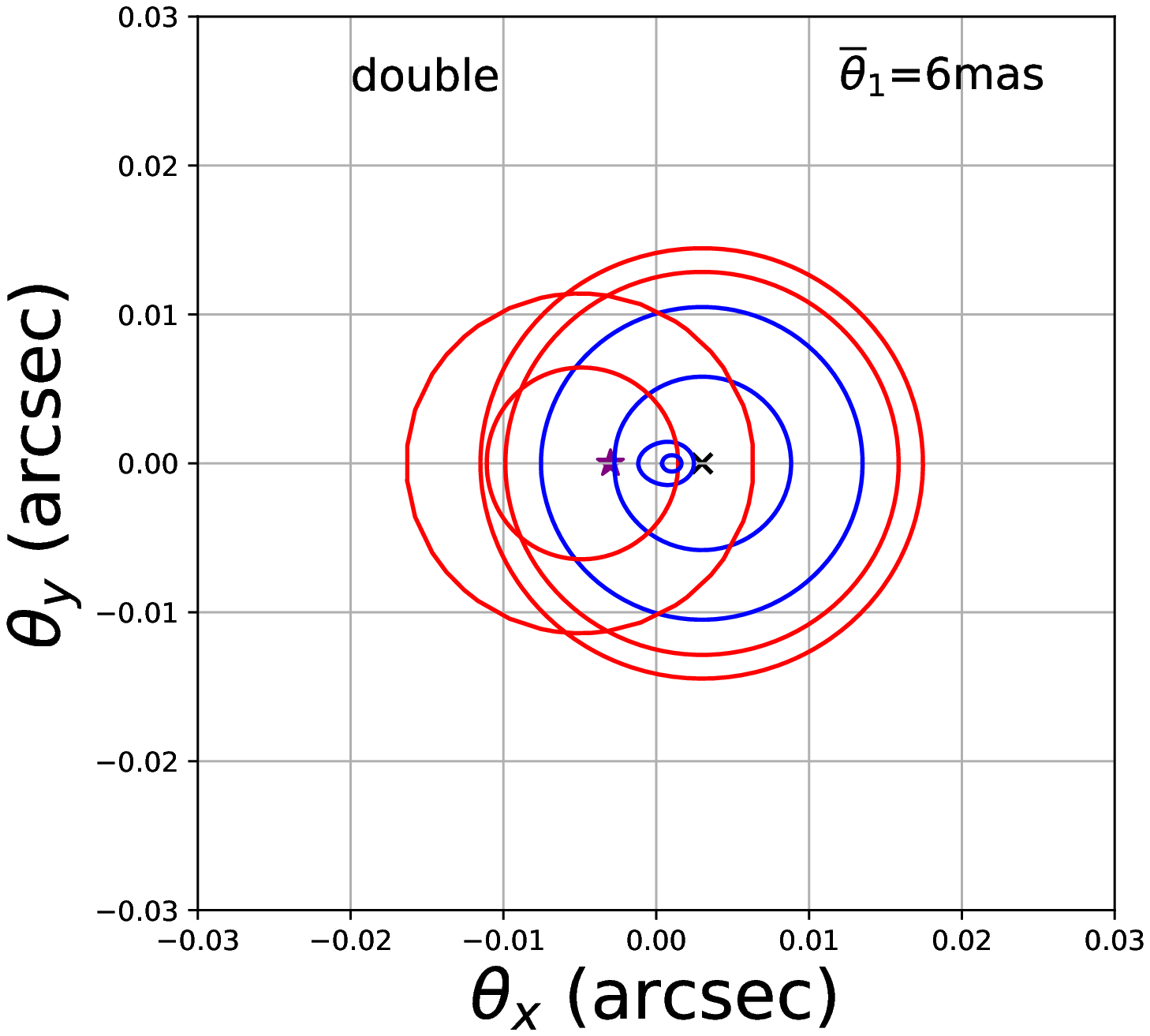}
\includegraphics[width=4cm]{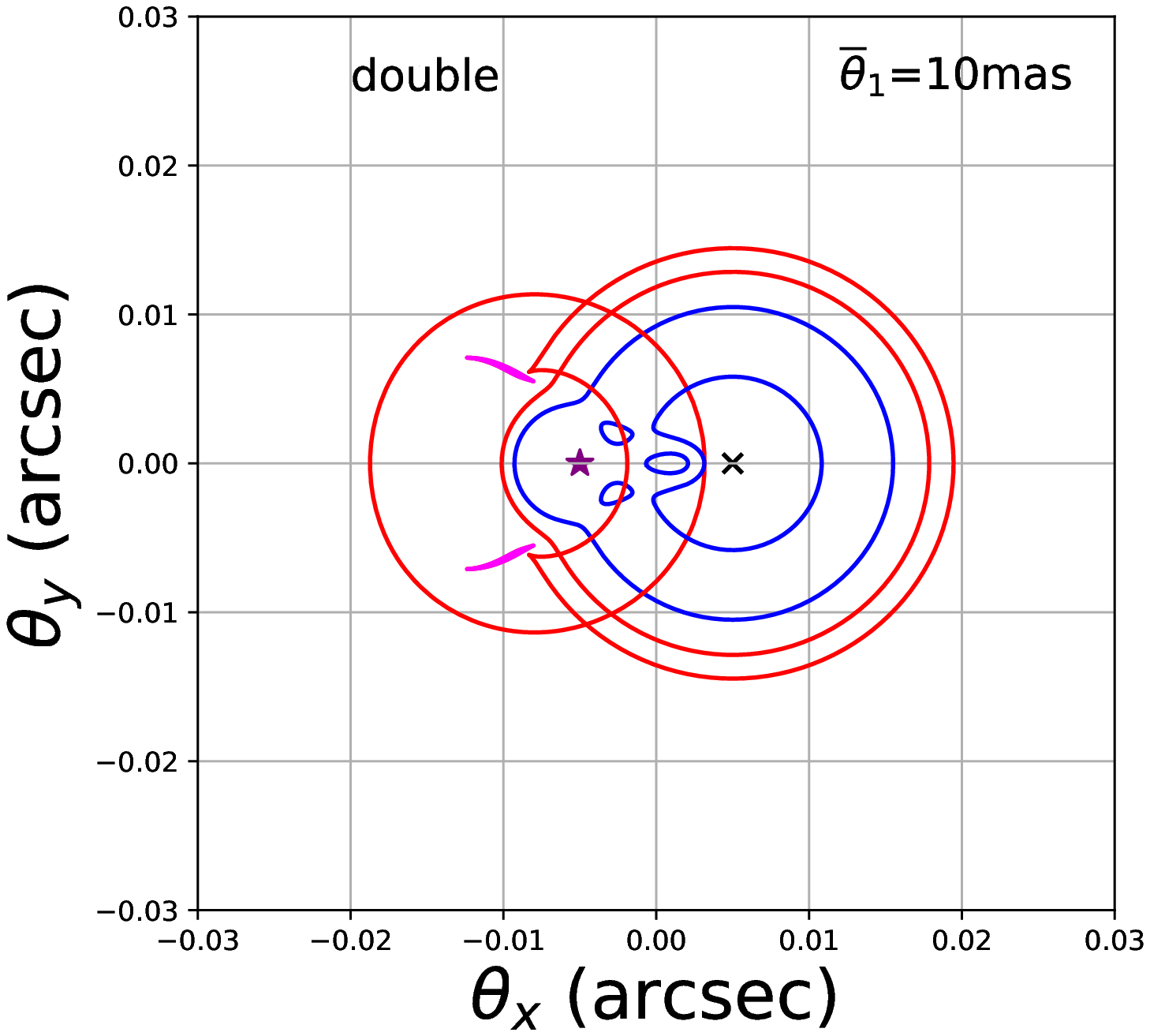}
\includegraphics[width=4cm]{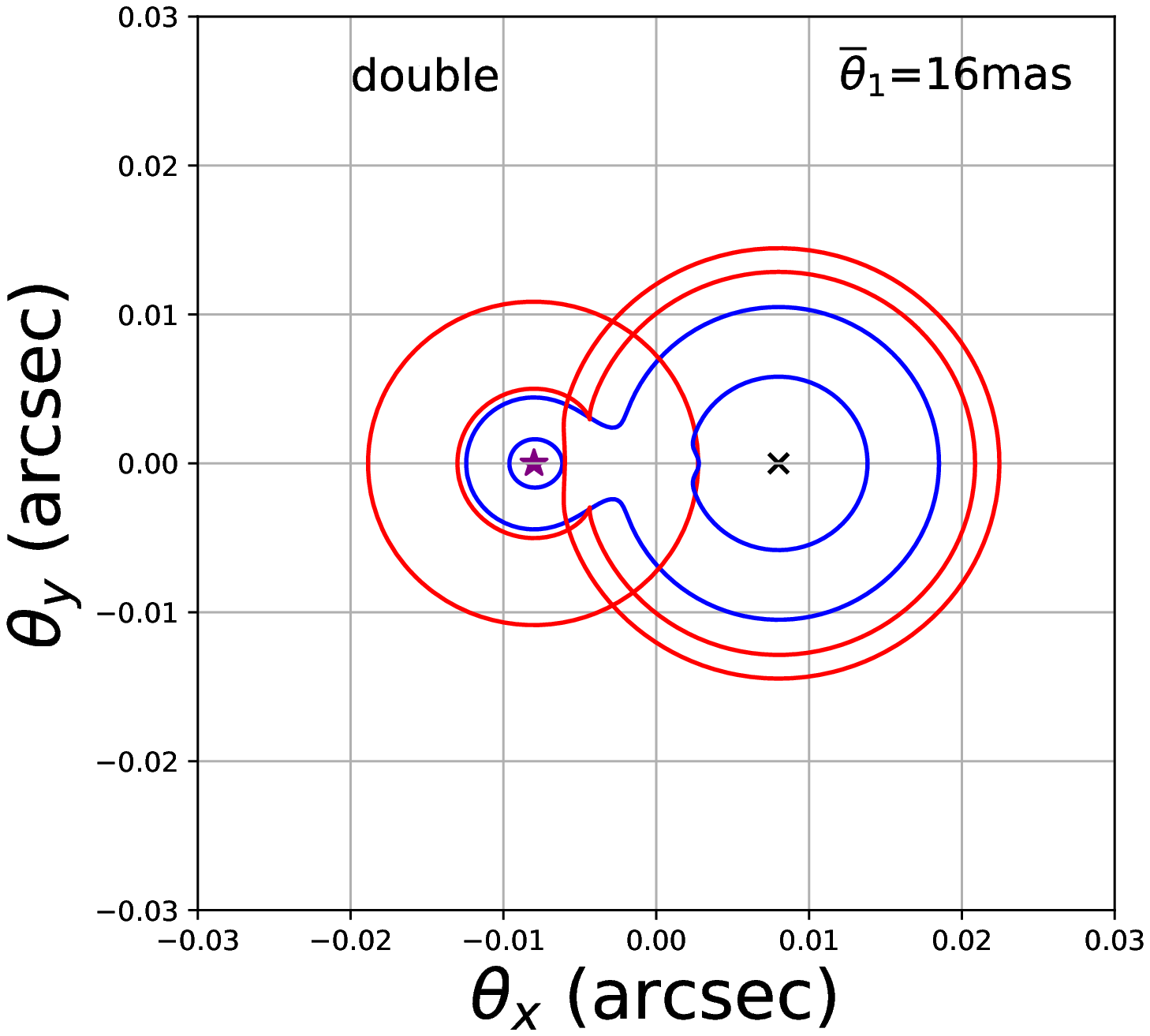}}
\centerline{
\includegraphics[width=4cm]{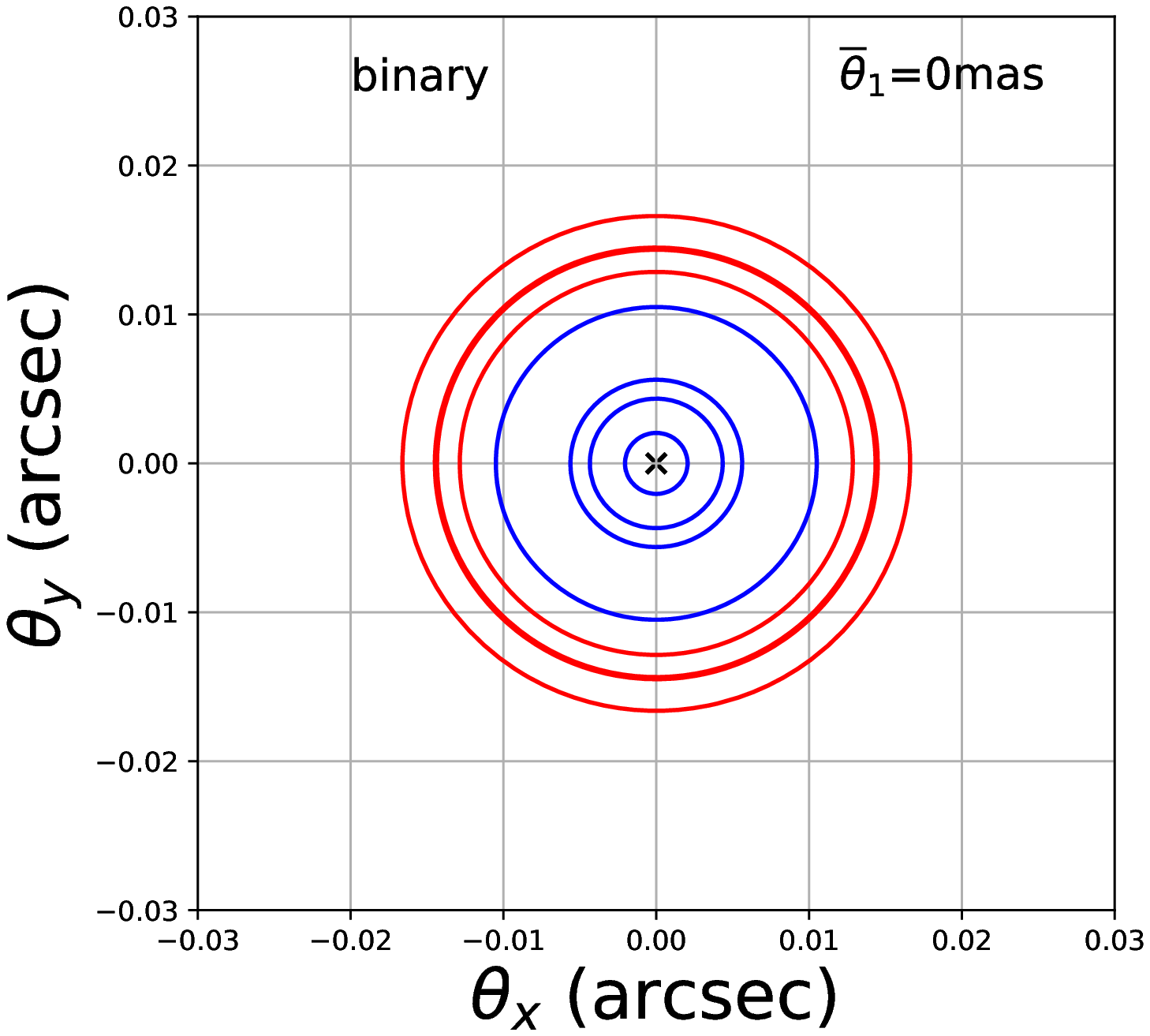}
\includegraphics[width=4cm]{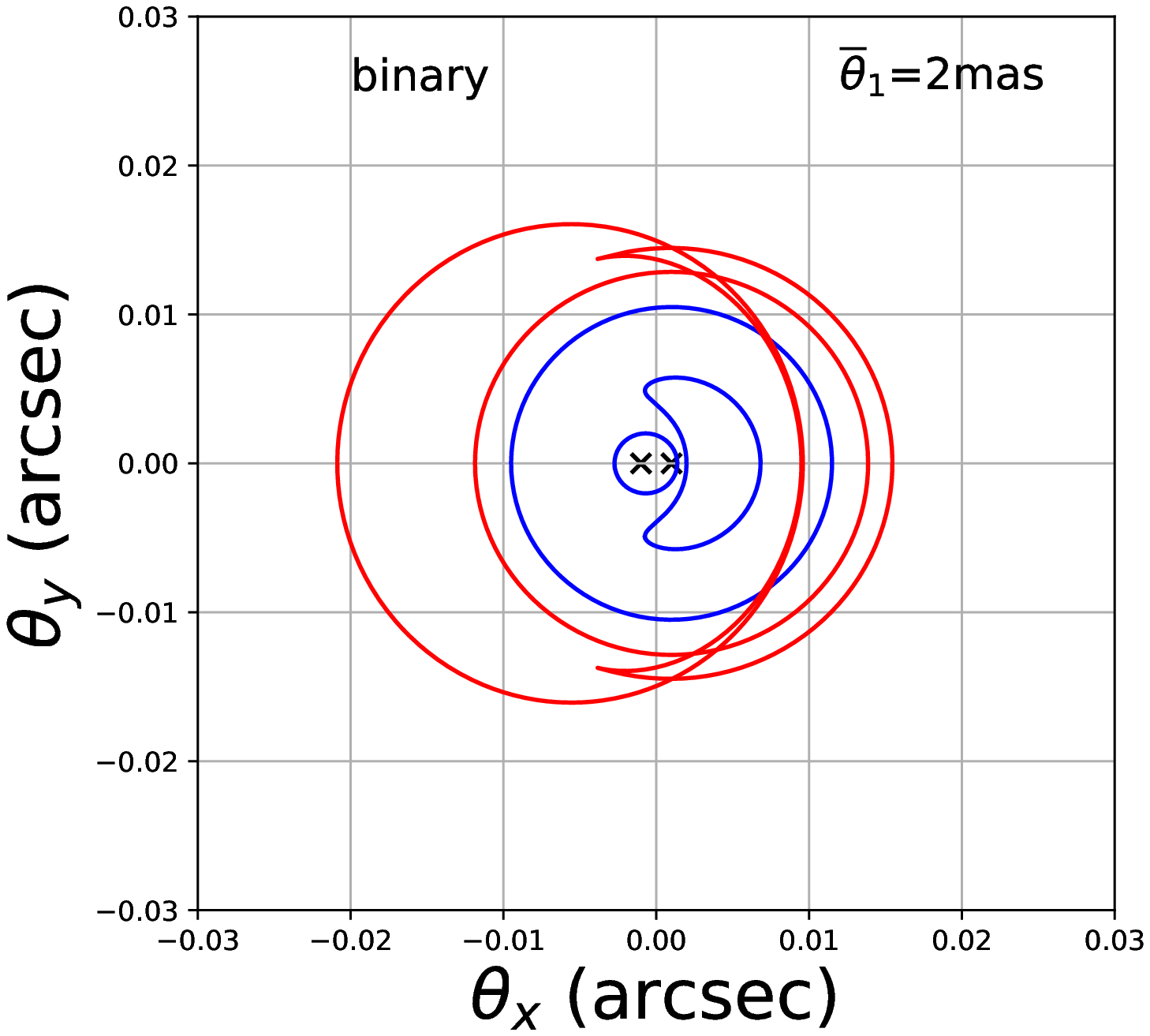}
\includegraphics[width=4cm]{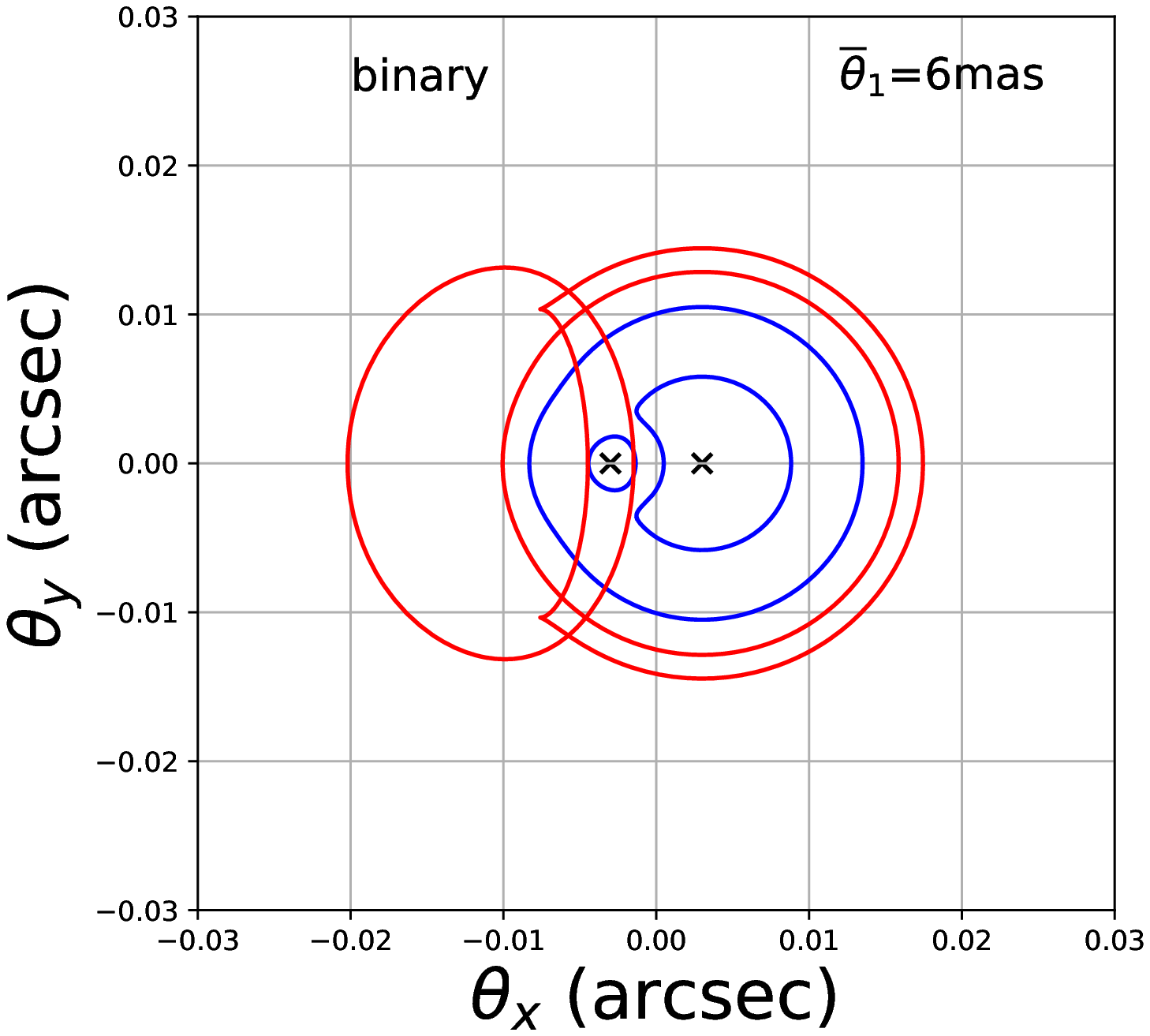}
\includegraphics[width=4cm]{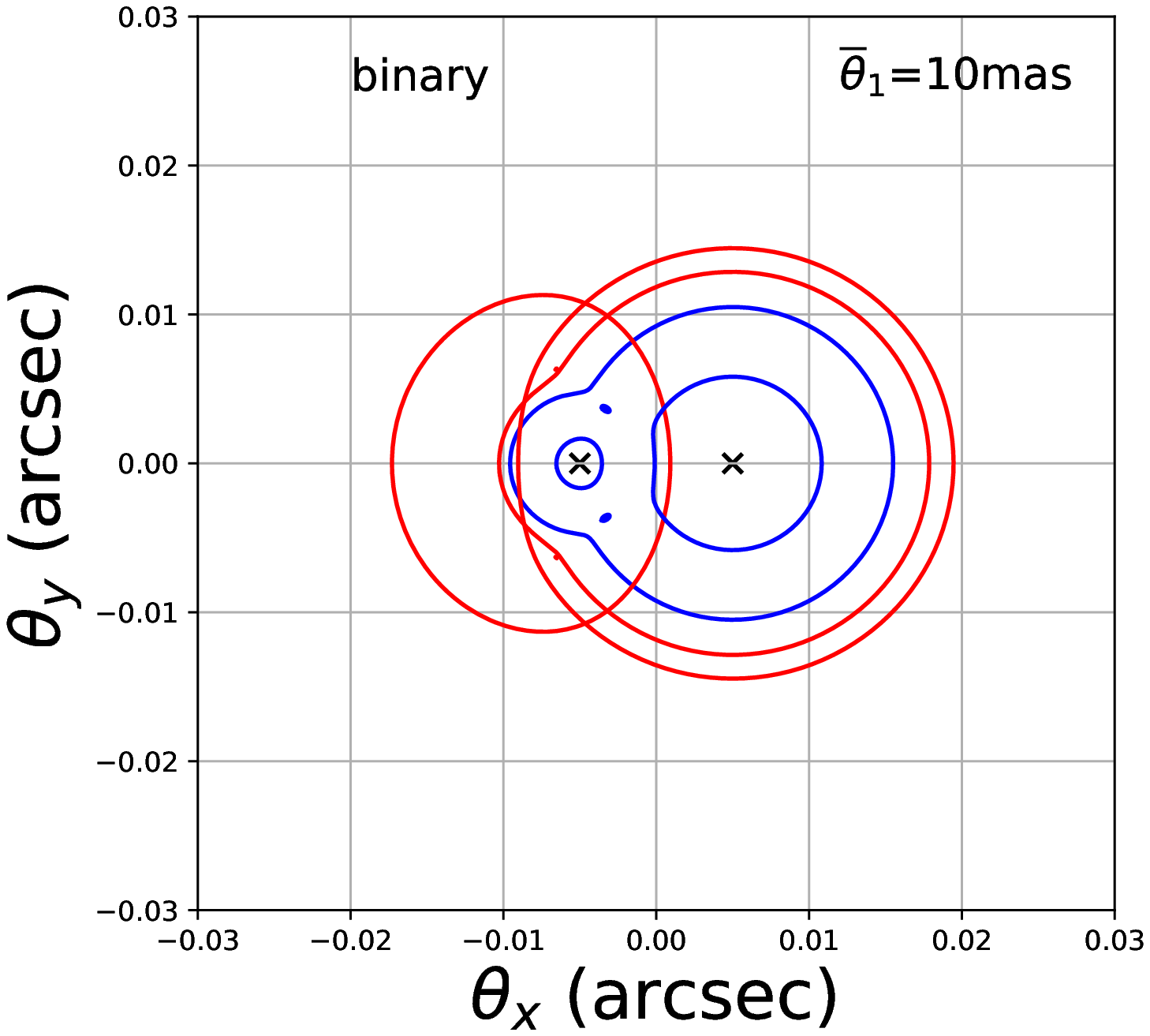}
\includegraphics[width=4cm]{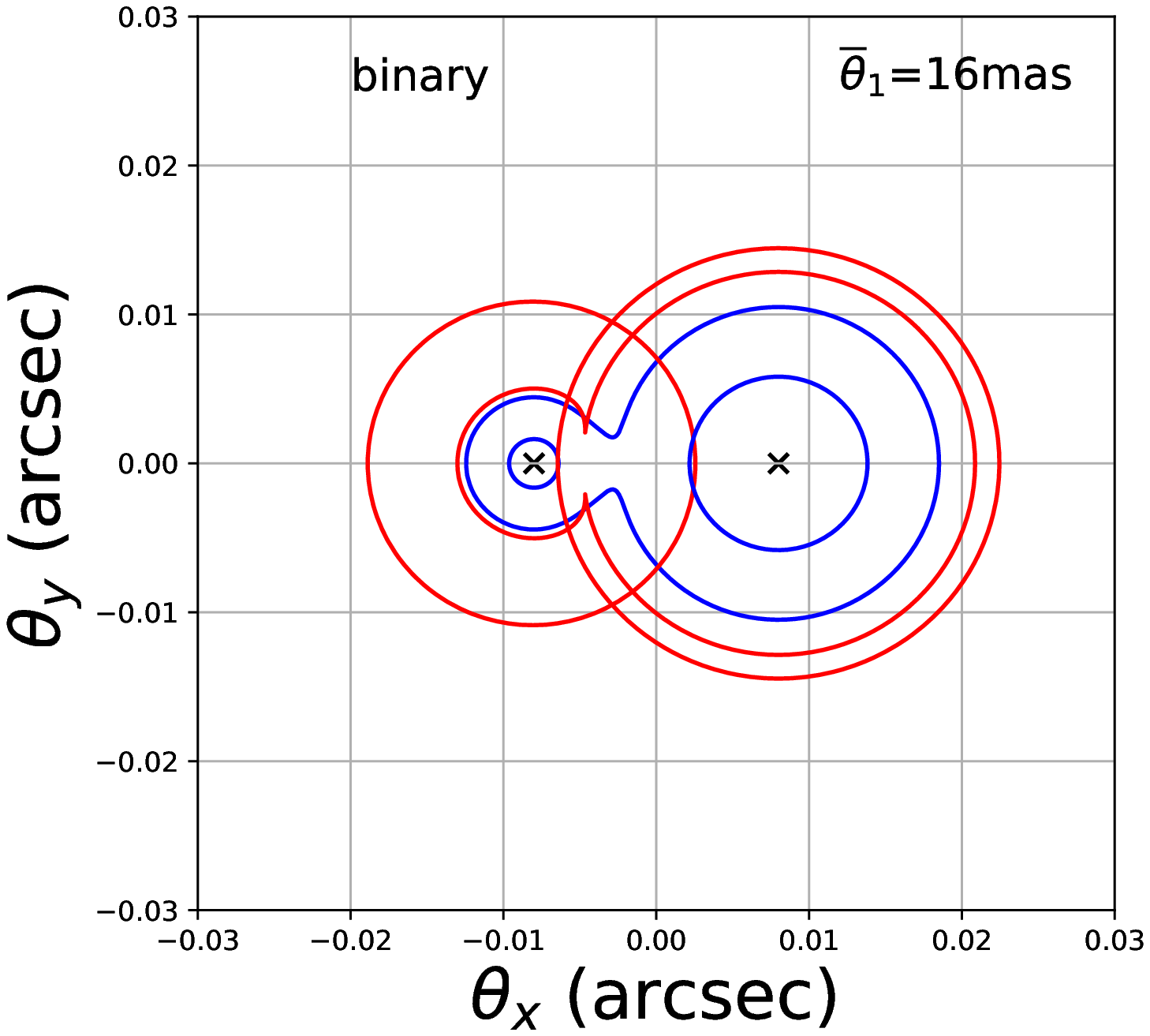}}
\caption{Critical curves (blue) and caustics (red) of a double lens with a potential given by Eq.~\ref{eq:double_potential} (the top panels), and binary lens with a potential given by Eq.~\ref{eq:binary_potential} (bottom panels). The redshift of the double lens are $z_d=0.0015$ (D1 lens, purple star),  $0.0005$ (D2 lens, black cross) respectively. The angular separation between the lenses (for both top and bottom) from left to right is for $\bar\theta_1=0,\,2,\, 6,\, 10,\, 16~\mbox{mas}$, respectively. }
\label{fig:cc-binary1}
\end{figure*}

We compare the lensing effects between a double lens in 3D, represented by the deflection potential of Eq.~\ref{eq:double_potential}, and a binary lens in 2D \citep{rogers&er19}, represented by the deflection potential of Eq.~\ref{eq:binary_potential}, to investigate the impact of alignment between lenses. 
In order to obtain comparable observables for both lensing configurations, we first calculate the two $\theta_0$ ($\theta_{01}$ and $\theta_{02}$) and $\sigma$ from a double lens, and adopt the same values in the binary lens. 
{For this case, the difference will come from the factor $x$ in the exponent of the first Gaussian lens in Eq.~\ref{eq:double_potential}.} 
One will see that such a factor has a small impact when the separation of the two lenses  $(\bar\theta_1)$ becomes large.
The electron number densities and lens {width of both Gaussians in} the double lens are the same for all the cases: $N_0=90$ pc\,cm$^{-3}$, and $\sigma=10^4$ AU. We first show the caustics and critical curves of the double lens in Fig.\,\ref{fig:cc-binary1}. We increase the angular separation between the two lenses from the left to the right panel. {The caustics and critical curves of the binary lens are shown in the bottom panels using Eq.\,\ref{eq:binary_potential}, they are almost identical to those generated by the double lens when the angular separation is sufficiently large, i.~e. $\bar\theta_1>\theta_0$.}
{In the fourth panel of double lens ($\bar{\theta_1}=10$ mas), the two small elongated caustics (shown in magenta) are not connected to the other curves. They are the corresponding caustics to the two small critical curves.}

\begin{figure*}
\centerline{\includegraphics[width=5.0cm]{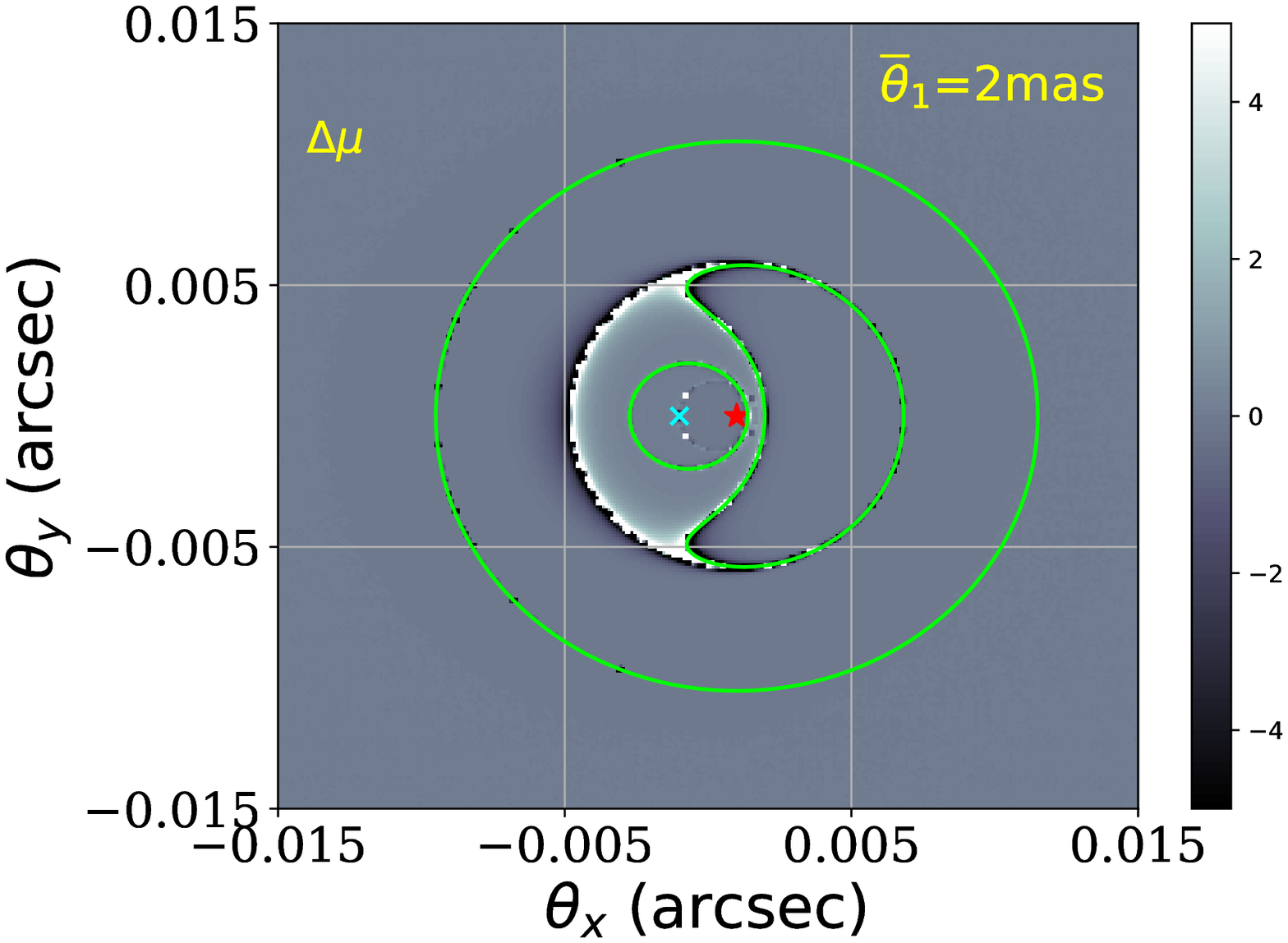}
\includegraphics[width=5.0cm]{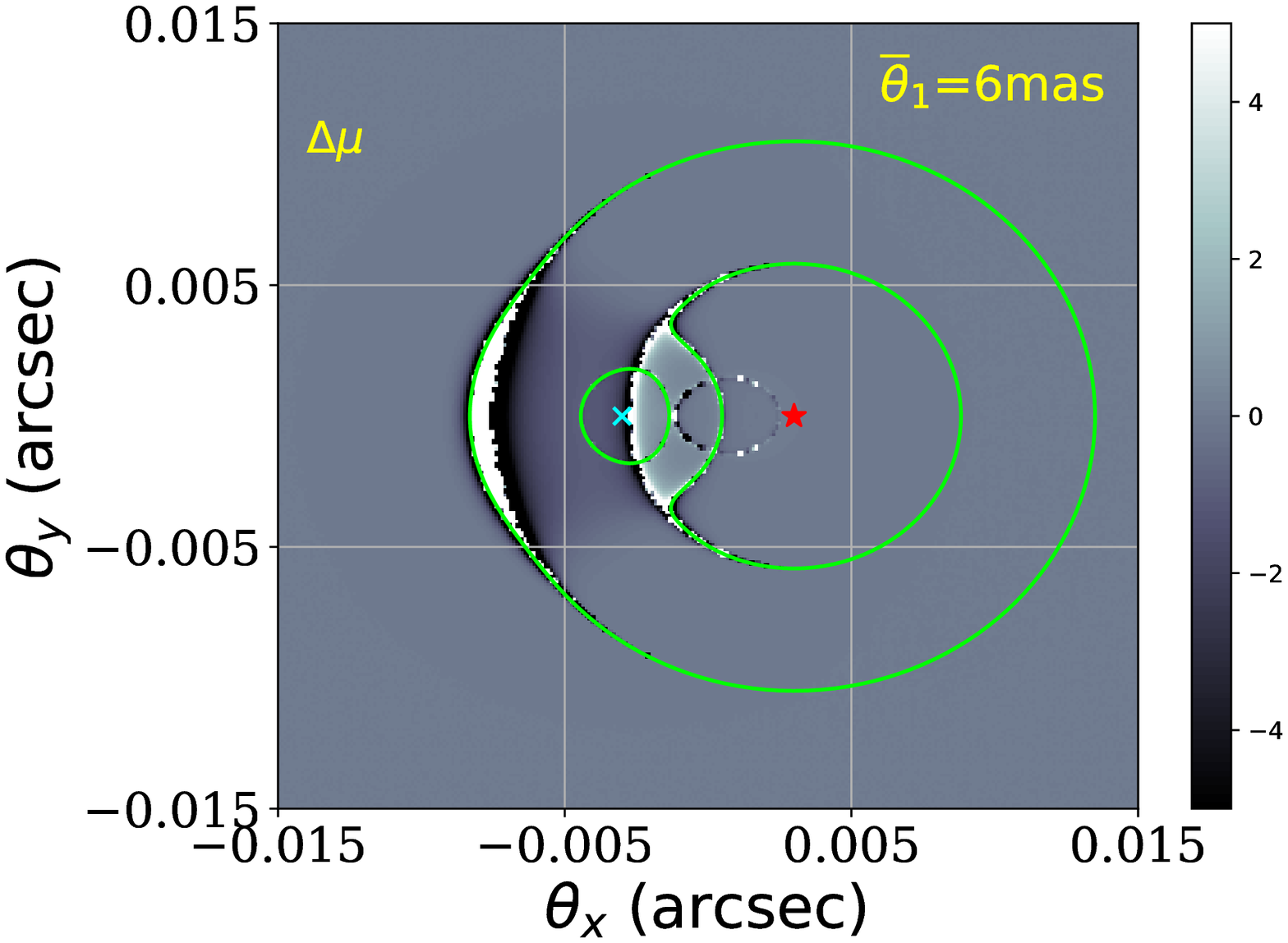}
\includegraphics[width=5.0cm]{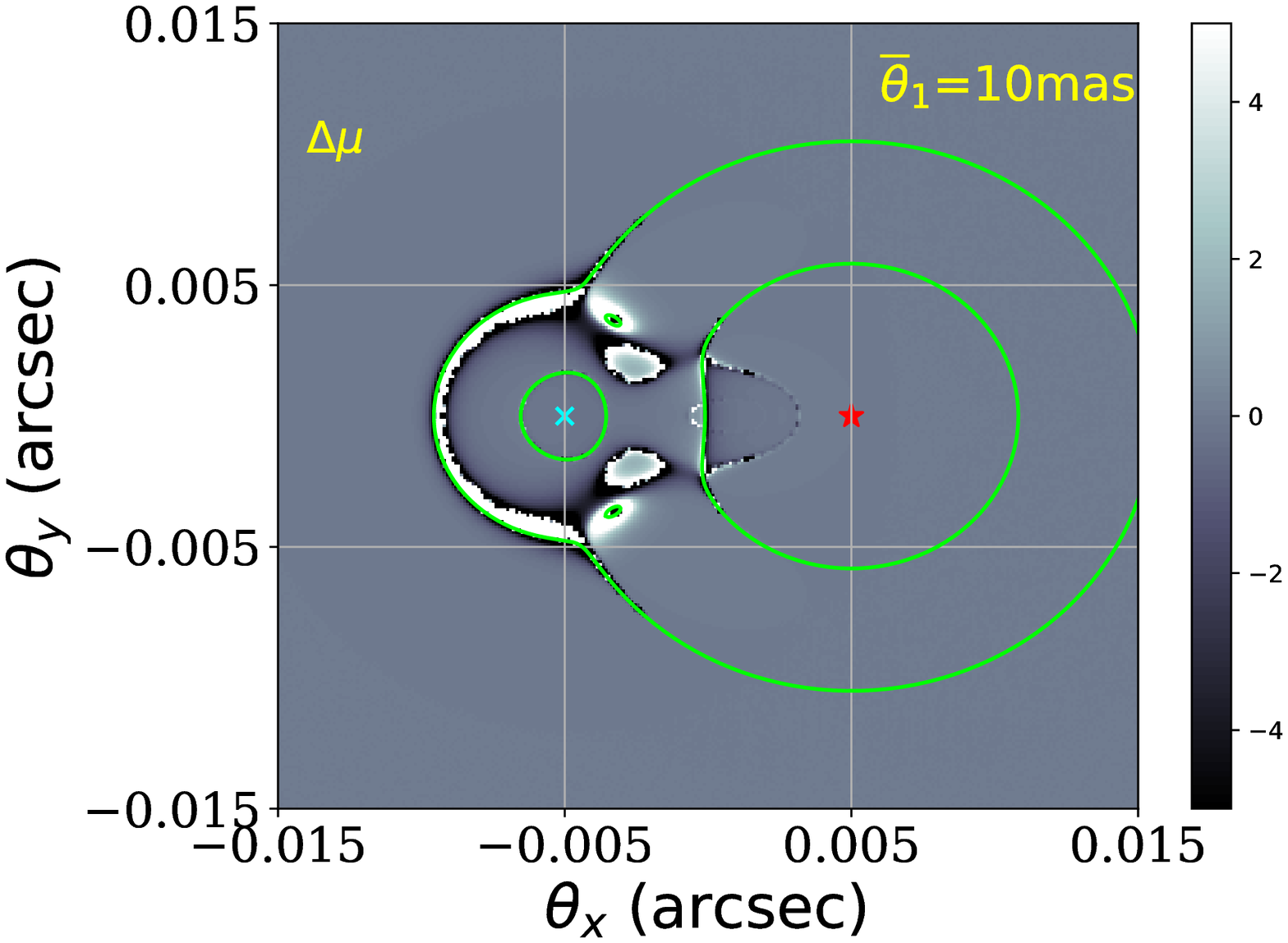}
\includegraphics[width=5.0cm]{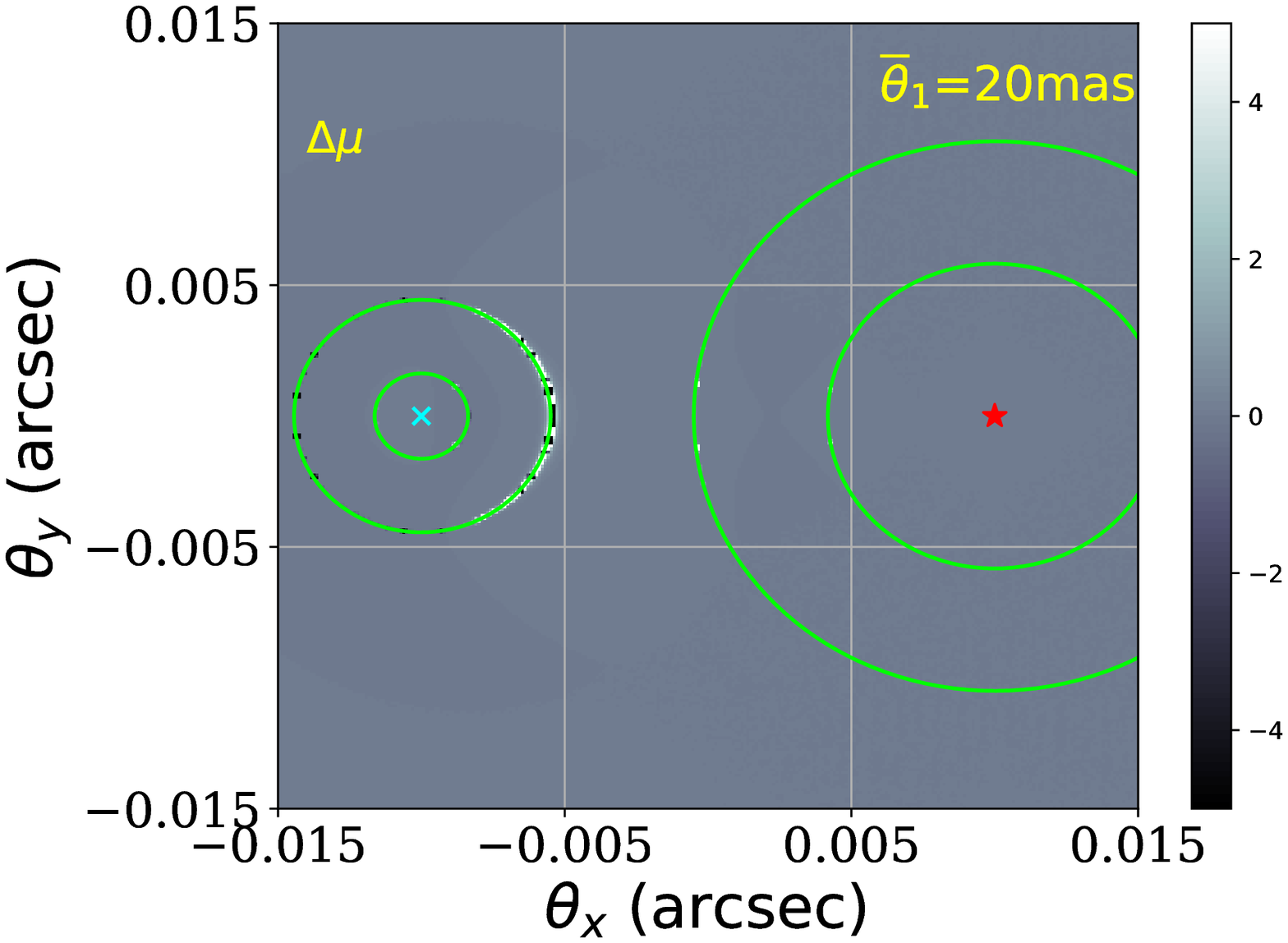}}
\centerline{\includegraphics[width=5.0cm]{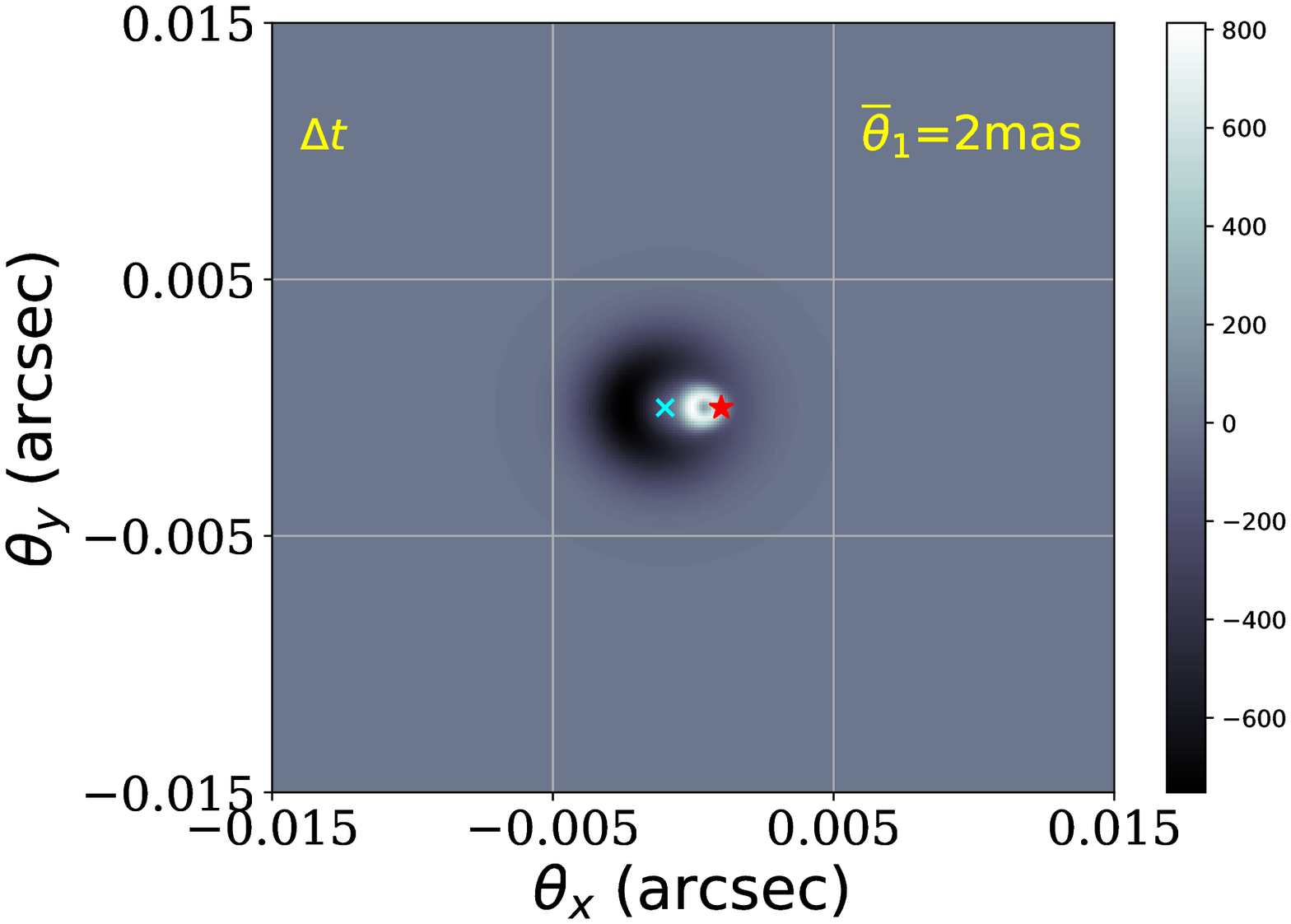}
\includegraphics[width=5.0cm]{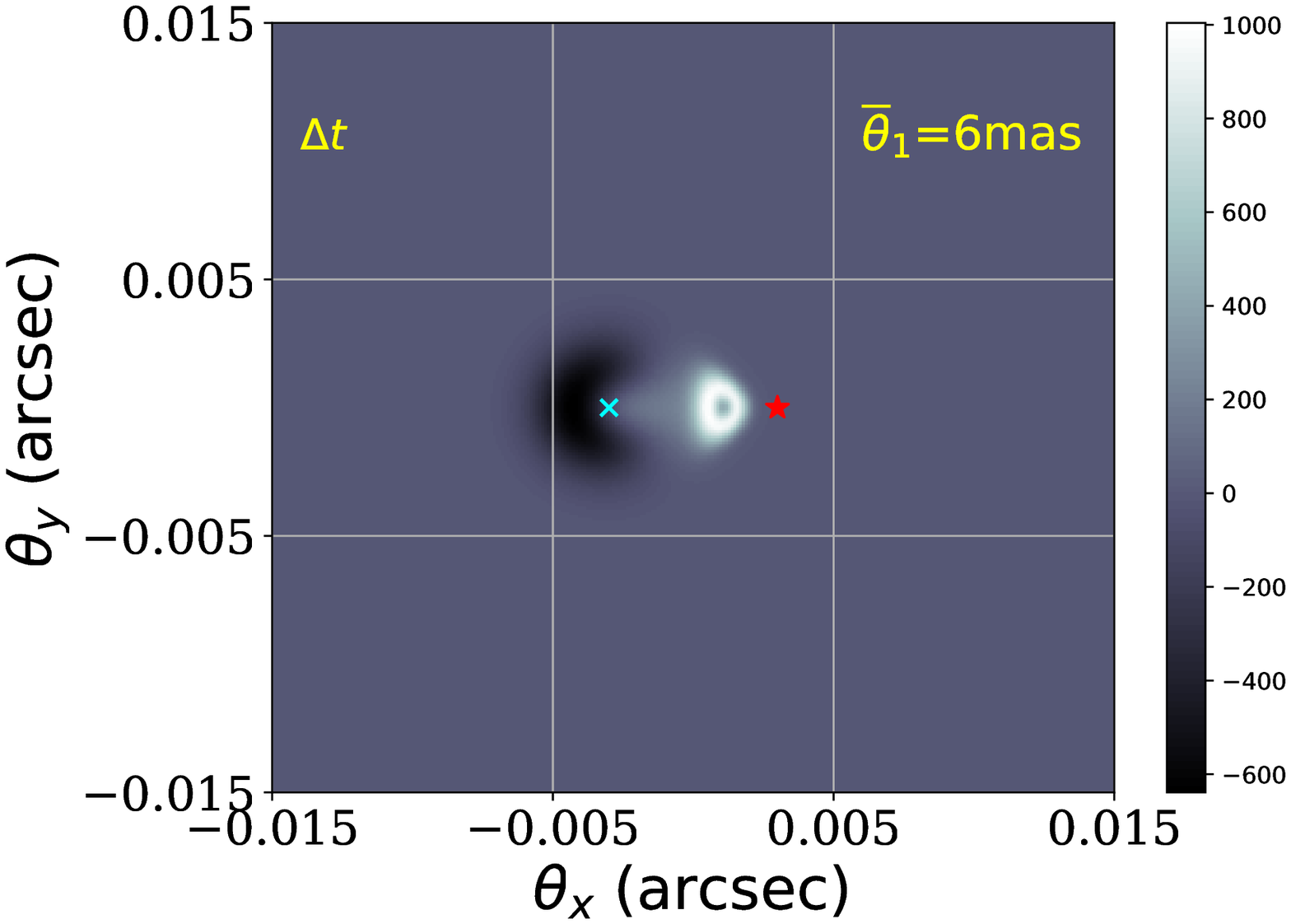}
\includegraphics[width=5.0cm]{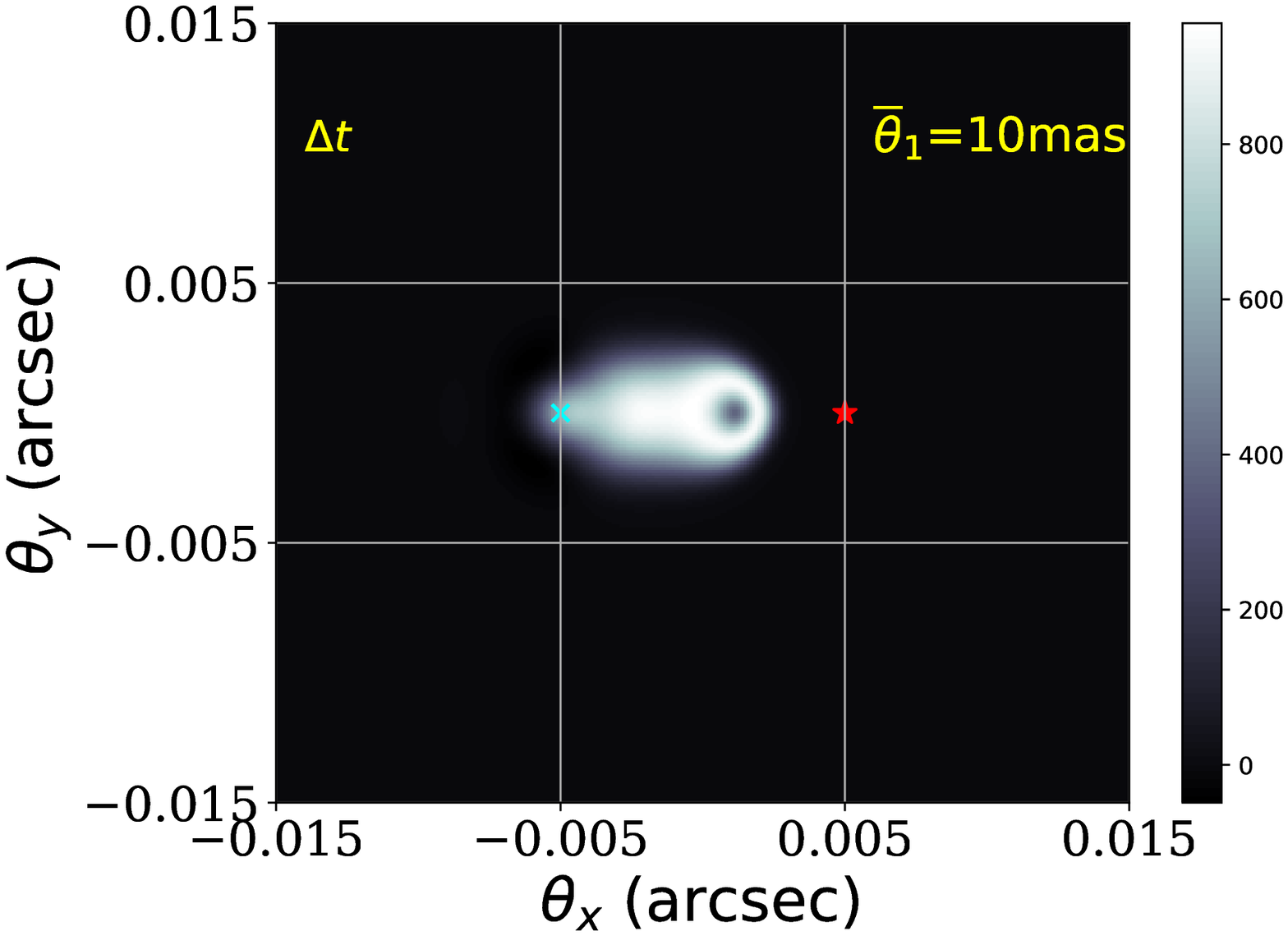}
\includegraphics[width=5.0cm]{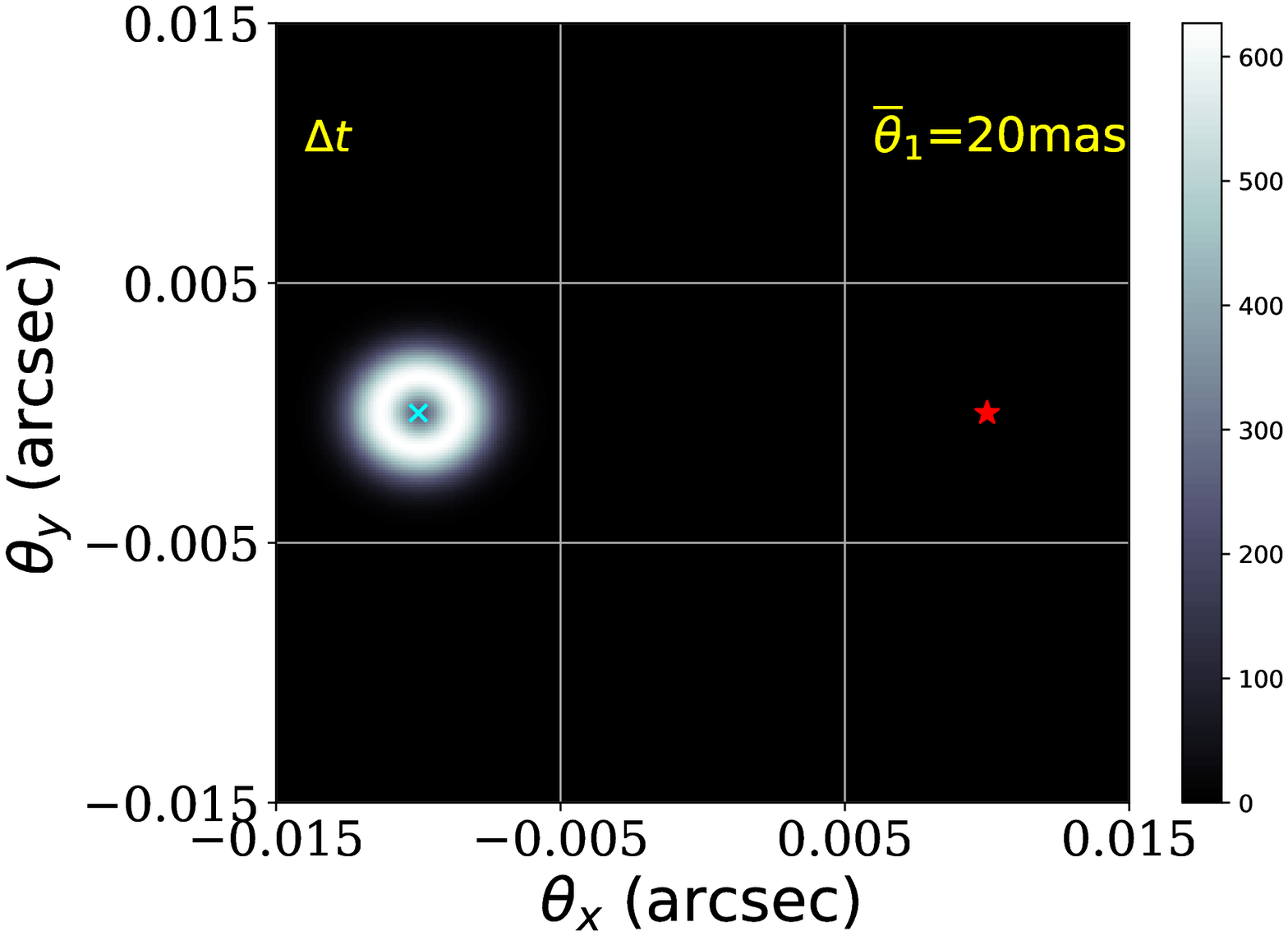}}
\caption{The difference of magnification (top) and time delay (bottom, in units of milli-seconds) between the double lens and binary lens: $(\mu_{\rm double}-\mu_{\rm binary})$, ($t_{\rm double}-t_{\rm binary}$). The green curves show the critical curves of the binary lens (top panels). The angular separation between the lens (for both top and bottom) from left to right is for $\bar\theta_1=2,\, 6,\, 10,\,  20~\mbox{mas}$, respectively. The cyan cross indicates the angular position of D1 lens ($z_d=0.0015$), and red star of D2 lens ($z_d=0.0005$). } 
\label{fig:rftimedelay}
\end{figure*}

We further compare the magnification and time delay maps between the binary lens and the double lens in Fig.\,\ref{fig:rftimedelay}. In order to calculate the time delay, the redshift for the binary lens is assumed to be the lens closest to the source in the double lens. Similar as that in caustics, the difference in magnification becomes small when the angular separation between the two lenses is sufficiently large. 
When the angular separation becomes small, there are significant differences in the magnification map. {The most significant differences in the magnification appear around the critical curves (green curves, representing the critical curves of the binary lens). As already discussed, wave effects can become significant in these areas, so that the two scenarios are hard to distinguish for observables that are located in the areas where geometrical optics is valid.}

\begin{figure}
\centerline{\includegraphics[width=8cm]{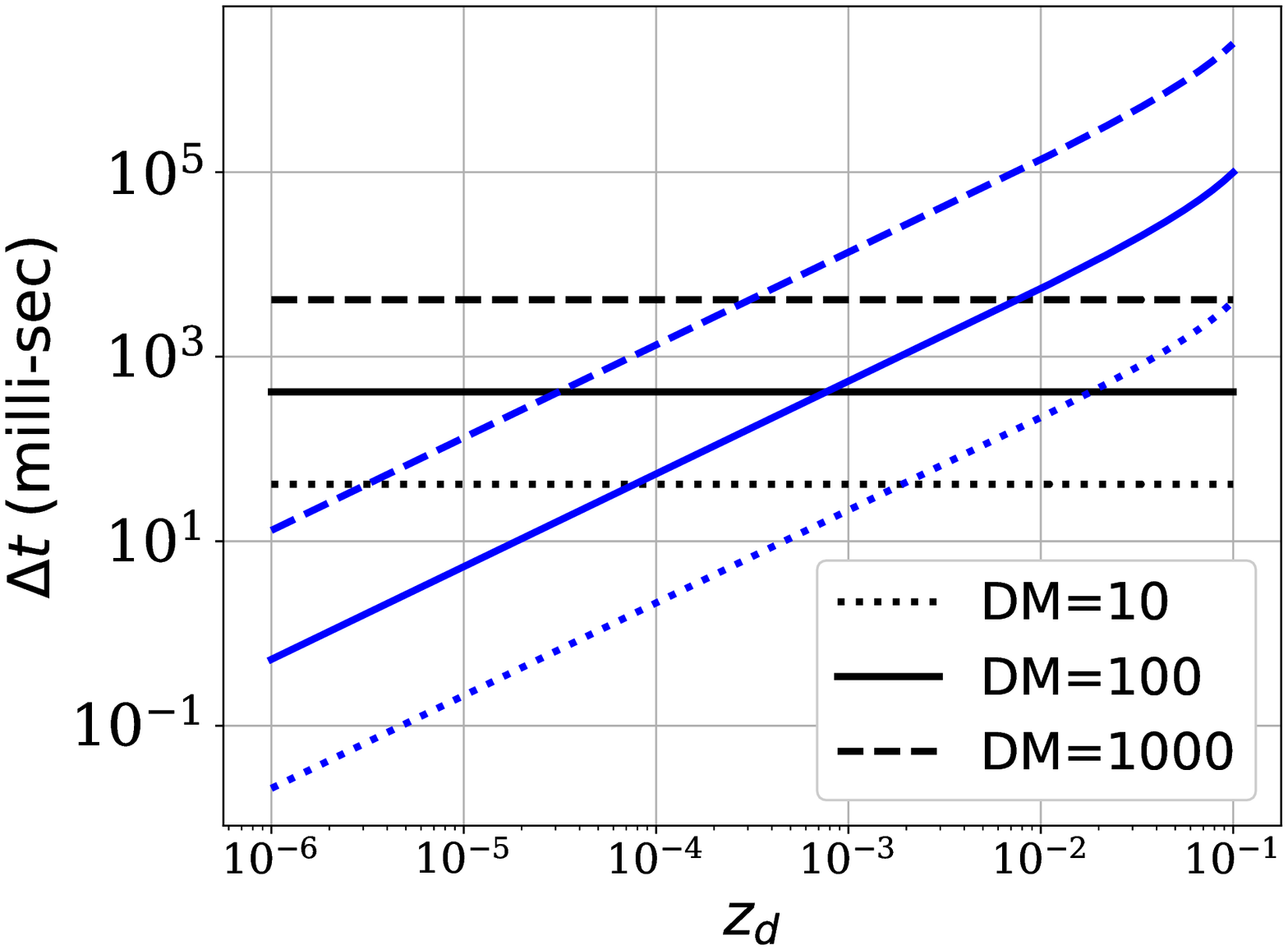}}  
\caption{Mock time delay as a function of lens redshift $z_d$. The source is placed at $z_s=0.19$. The black (blue) lines show the dispersive (geometric) time delay. The dotted, solid, and dashed line show the delay for DM=10, 100, 1000 pc\,cm$^{-3}$. In the geometric delay, constant deflection angle of $\alpha=$2, 10, 50~mas are used. 
}
\label{fig:timedelay}
\end{figure}

{The time delay, however, will show significant differences (bottom panels in Fig.\,\ref{fig:rftimedelay}). The time delay difference appears at different spatial positions than the differences in magnification and thus helps to distinguish the binary from the double lens.}
With a time resolution of milli-seconds in radio observations today, these differences can be easily observed. 
In particular, the geometric delay has an extra dependence on the distance to the lens, while the dispersive delay is weakly dependent on the redshift of the lens or the source. We compare the dispersive delay and geometric delay for lenses at different redshift in Fig.\,\ref{fig:timedelay}. For a given deflection angle, the geometric part will dominate the total time delay when the lens is located at a sufficiently high redshift. Thus, the time delay provides strong constraints on the distance to the lens and time delay differences can distinguish between single- and multi-plane lensing configurations.


\subsection{Time delays of a $\delta-$function profile and a Gaussian profile}
\label{sec:timedelay}
\begin{figure}
	\includegraphics[width=8cm]{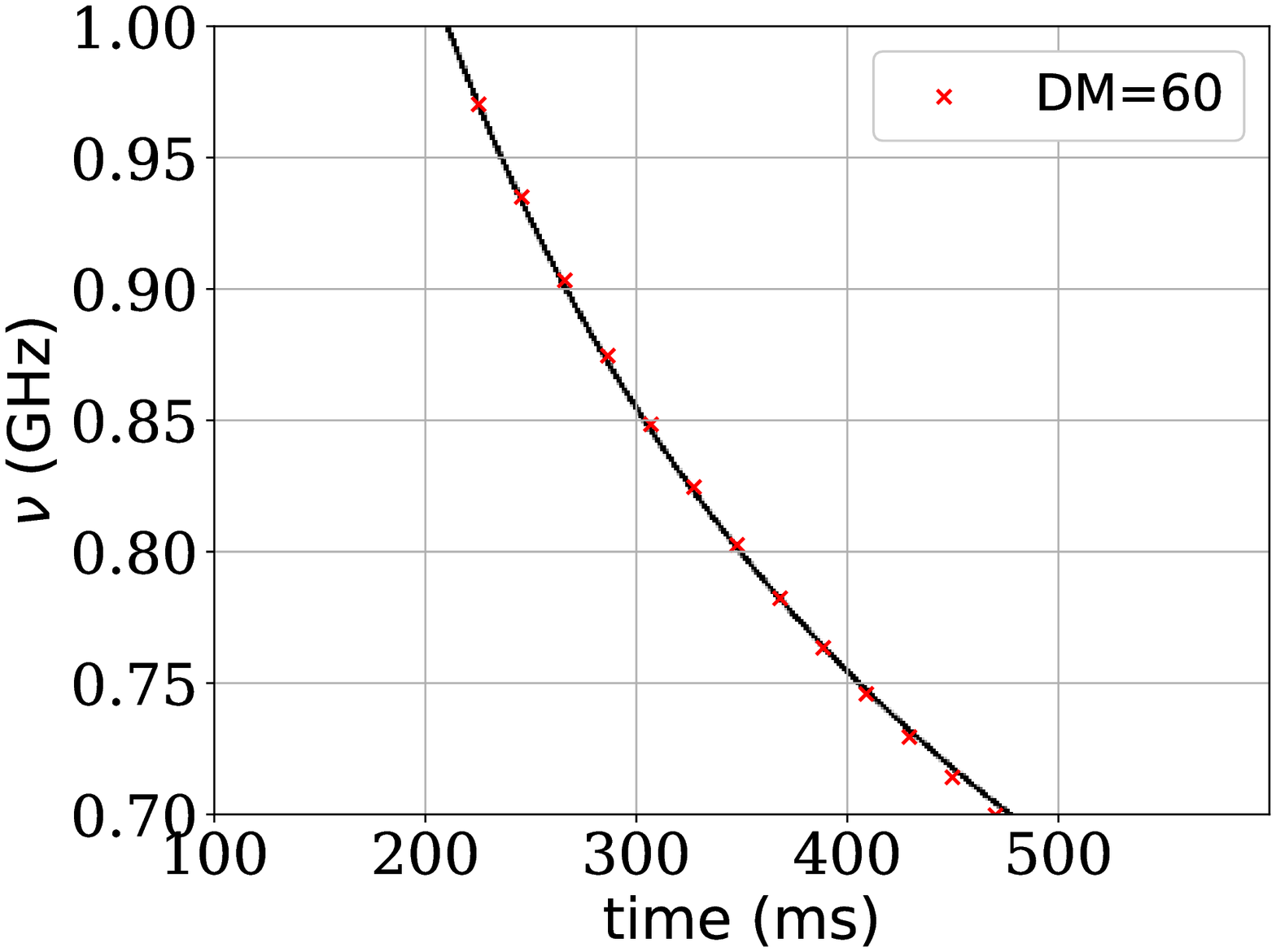}
	\caption{Simulated radio dispersion signal. The grey shadow presents the time delay signal of double lens (Table\,\ref{tab:single_vs_dB}, $\beta=10$ mas). The red crosses show the analytical curves of Eq.\,\ref{eq:theoryDM} with DM=$60$ pc\,cm$^{-3}$, and $t_0=40$ ms. Only the brightest image is shown in this figure.}
	\label{fig:single-spec}
\end{figure}
The time delay in plasma lensing is frequency dependent. The time delay-frequency relation of a compact pulse radio source is frequently used to estimate the DM \citep[e.g.][]{petroff2016}. The theoretical prediction is given by
\be
t(\nu)-t_0 \approx 4.16 {\rm ms} \rund{\dfrac{\rm DM}{\nu^2}},
\label{eq:theoryDM}
\ee
where the DM will always be given in units of pc\,cm$^{-3}$, $\nu$ is given in units of GHz. It has been noticed that there exist several biases in the estimate of electron density from DM \citep{2020arXiv200702886K}. Moreover, the geometric delay by lensing is non-negligible when the deflection is large \citep{2020CQGra..37t5017T}, and will further deviate from the relation of the frequency-time delay from Eq.\,\ref{eq:theoryDM}, \citep[e.g.][]{2016ApJ...817...16C,er+2020}.
In Fig.\,\ref{fig:single-spec}, we compare the double lensing scenarios and calculate the frequency over time delay for the radio signal for frequencies between $0.7-1$ GHz. The DM relations according to Eq.\,\ref{eq:theoryDM} are presented as well. The single lens and single D2 cases are not shown as the time delays are too small to be seen. It is possible to find an electron density (constant DM) to generate a similar curve using Eq.\,\ref{eq:theoryDM}, although the density will be different from those of the lenses (higher in our case).

\begin{figure}
	\includegraphics[width=8cm]{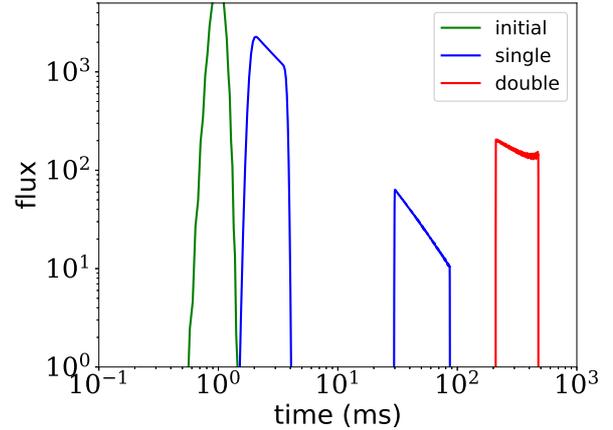}
	\includegraphics[width=8cm]{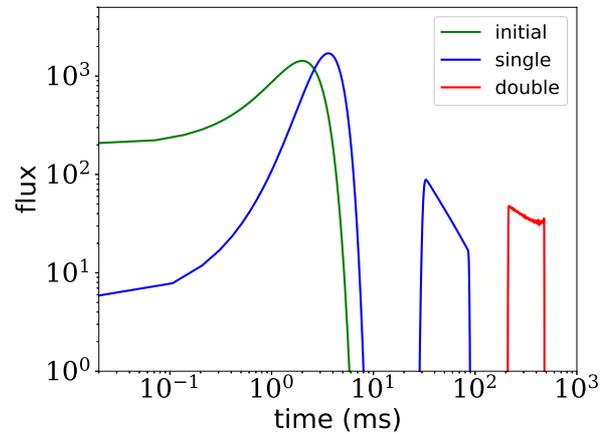}
	\caption{Simulated flux of radio signals over the band 0.7 to 1 GHz. The unit of the flux is arbitrary. $Top-$ for a $\delta-$function intrinsic shape; $bottom-$ for a Gaussian intrinsic shape. The source position of single lens is $\beta=8$ mas.}
	\label{fig:single-flux}
\end{figure}

We also calculate the simulated total flux of a pulsar signal. The initial spectrum of the source is assumed to be a sharp pulse
\be
f(\nu,t_0) \propto \delta(t-t_0)/\nu^2,
\ee
or, alternatively we consider a Gaussian signal
\be
f(\nu,t_0) \propto {\rm exp}\rund{-\frac{(t-t_0)^2}{2\sigma^2_{pw}}} /\nu^2,
\ee
where $\sigma_{pw}$ is the width of the signal. We integrate the flux from 0.7 to 1 GHz, taking into account the time delay and the magnification by lensing,
\be
F(t)=\int d\nu f(\nu,t-t_d(\nu,z_d),z_s)\,\mu(\nu),
\ee
where $t_d(\nu)$ is the total time delay induced by plasma lensing. The single lens and double lenses are shown in Fig.\,\ref{fig:single-flux}. The flux is plotted in arbitrary units. In the single lens case, we choose the source position to be $\beta=8$ mas. 
There exist multiple images with different magnifications and time delays. In both scenarios, the second image has a slight de-magnification. Thus a second pulse with significantly longer time delay can be seen, see e.g. the blue curves for a single-plane lens. In the double lens, the time delay between the two images has a relatively small difference (209 ms and 341 ms). Thus the fluxes from the two images blend, which can be seen from the second rise in the red curves. The single-lens scenario (blue curves) can be easily distinguished from the double lens (red curves), even if the multiple images are not spatially resolved. The degeneracy in the image position and magnification between the single lens and the double lens (Table\,\ref{tab:single_vs_dB}), can be easily broken in Fig.\,\ref{fig:single-flux}. 

As an additional comparison, we adopt a Gaussian for the intrinsic shape of the pulse with width $\sigma_{pw}=1~\mbox{mas}$. If the lens causes a small time delay, one can still reproduce the intrinsic shape of the pulse, e.g. the blue curves. Otherwise, the pulse shape will be dominated by the lensing properties. For instance, the red curves from the two panels will be difficult to distinguish.

\section{Conclusions}
\label{sec:conclusions}
In this work, we study the effects of multiple lens planes in plasma lensing. Plasma lensing shares the same formalism with gravitational lensing, but plasma lenses also have significant dissimilarities. In gravitational lensing, the mass distribution is concentrated and can be modelled from dynamics or numerical simulations. Usually, the deflections are mainly caused by a massive structure at one redshift. In plasma lensing, the constraints on the electron distribution are currently weak, and the extension of the electron density along the line of sight can be extremely large. We investigate when a single plasma lens can be fitted to the data and when an extension to multiple lens planes is required in this work. 

Our analysis focuses on axisymmetric Gaussian lenses because they can also be superposed to form more complex deflection patterns, if needed. We compare the single-plane lens with double-plane lenses and find that an effective single plane lens can mimic some image properties generated by multi-plane lensing, but not all of them at the same time. For example, we can set up a double lens which produces image positions and magnifications similar to those of a single lens, but then, the time delays show significant differences (see Table~\ref{tab:single_vs_dB}). 
We also compare double lenses at two redshifts with binary lenses at one redshift at intermediate distances between the source and the observer. When the lens parameters to characterise the two Gaussians, $\theta_0,\sigma$, are the same, the magnification maps produced by the two scenarios show differences which are difficult to identify. But, again, their time delay is sufficiently different. 
We show that the time-delay-frequency relation differs from the theoretical prediction when a constant DM is adopted. 
Although constraining plasma in several lens planes, especially finding the distances to the planes, is difficult having only a few observables, the different delay curves can provide a potential way to infer structures of the electron distribution along the line of sight, as it is already implemented for observations e.g. in \cite{2019ascl.soft10004S}. 
We also show that one can mimic the time delay-frequency relation of a double-plane lens using a single plane lens, but a bias will arise in the estimate of electron density. Thus, besides the several uncertainties pointed by \citet{2020arXiv200702886K}, another bias exists due to the degeneracy of single-plane and multi-plane lens in Eq.\,\ref{eq:theoryDM}. For a precise estimate of the electron density, extra factors need to be taken into account, e.g. the density gradient.

Additional constraints or assumptions may still be necessary as we lack knowledge about the intrinsic pulse shape of the source. The pulse width of the source can be broadened significantly by the plasma dispersion, if the intrinsic width is small. The overall shape can be changed as well due to the frequency-dependent time delay and the (de-)magnification effects.
In addition, we mainly show the time delay-frequency and flux relation for the brightest image, which is the image closest to the source position in all the cases we studied in this work. In several cases, the secondary image, which is mildly de-magnified can cause another peak in the flux spectrum. The secondary image provides additional constraints which can be used to distinguish different lensing scenarios.

The density profile we adopted in this work is a Gaussian function. Although other profiles, such as a power law or an exponential one exist as well, we focus on the Gaussian profile for our plasma lenses as it is an analogous to a point mass model in gravitational lensing. Hence, the Gaussian plasma lens can be considered as a building block to generate more general profiles. Further non-axisymmetrical distribution studies for structures along the line of sight, e.g. more lens planes, are left for future work. In a preliminary study, we find plasma lenses in the Milky Way may dominate some lensing effects, and again the time delay information is necessary to constrain the lens distance. Modelling a realistic plasma lens profile along the line of sight may require further inputs from observations and simulations. For multiple planes with high electron densities, we find that the lenses have the possibility of generating a series of images, as after every lens plane the number of the images increases. Such phenomena can be interesting evidence for the complex structure of plasma lensing along the line of sight. 
Moreover, in plasma lensing the deflection depends on the wavelength of the photon at the lens redshift. For the lensing system at a cosmological distance, the difference of wavelength between that at source, lens and observer will become significant. Such a difference can affect the estimate of the electron density of the plasma lens and has to be included in the future study. 

\section*{Acknowledgements}
We thank the referee for valuable constructive comments to the manuscript. We also thank Adam Rogers for comments on the draft. XE is supported by NSFC Grant No. 11873006. SM is supported by the National Key Research and Development Program of China No. 2018YFA0404501, and NSFC Grant No. 11821303, 11761131004 and 11761141012, SM also acknowledges the SWIFAR visiting fellow program under which he had a fruitful visit to the South-Western Institute for Astronomy Research, Yunnan University.
\section*{Data availability}
The data underlying this article will be shared on reasonable request to the corresponding author.
\bibliographystyle{mnras}
\bibliography{lens.bib,refbooks.bib,stronglens.bib,plasmalens.bib,frb.bib,wavepaper.bib,multi.bib,survey.bib,shear.bib,refcos.bib,galaxy.bib}

\end{document}